\documentclass[12pt,epsf]{article}
\usepackage{amssymb,amsmath}
\usepackage{graphicx}
\usepackage[dvips]{color}

\newcommand{\be}{\begin{equation}}
\newcommand{\ee}{\end{equation}}
\newcommand{\bea}{\begin{eqnarray}}
\newcommand{\eea}{\end{eqnarray}}
\newcommand{\beas}{\begin{eqnarray*}}
\newcommand{\eeas}{\end{eqnarray*}}
\newcommand{\ba}{\begin{array}}
\newcommand{\ea}{\end{array}}

\newcommand{\nbox}{{\,\lower0.9pt\vbox{\hrule \hbox{\vrule height 0.2 cm \hskip 0.19 cm \vrule height 0.2 cm}\hrule}\,}}

\def\href#1#2{#2}
\input{tcilatex}

\begin{document}

\begin{titlepage}
\hfill
\vbox{
    \halign{#\hfil         \cr
           } 
      }  

\hbox to \hsize{{}\hss \vtop{ \hbox{}

}}

\vspace*{20mm}
\begin{center}
{\Large \bf Strings on Bubbling Geometries\\}

\vspace*{15mm} \vspace*{1mm} Hai Lin{\footnote{e-mail: hai.lin@usc.es}}, Alexander Morisse{\footnote
{e-mail: alexander.morisse@usc.es}} and Jonathan P. Shock{\footnote {e-mail: shock@fpaxp1.usc.es}}

\vspace*{1cm}

{\it Department of Particle Physics, Faculty of Physics, \\
University of Santiago de Compostela, 15782, Santiago de Compostela, Spain \\}

\vspace*{1cm}
\end{center}

\begin{abstract}
We study gauge theory operators which take the form of a product
of a trace with a Schur polynomial, and their string theory duals.
These states represent strings excited on bubbling AdS geometries
which are dual to the Schur polynomials. These geometries
generically take the form of multiple annuli in the phase
space plane. We study the coherent state wavefunction of the lattice, which
labels the trace part of the operator, for a general Young tableau
and their dual description on the droplet plane with a general
concentric ring pattern. In addition we identify a density matrix
over the coherent states on all the geometries within a fixed
constraint. This density matrix may be used to calculate the
entropy of a given ensemble of operators. We finally recover the
BMN string spectrum along the geodesic near any circle from the
ansatz of the coherent state wavefunction.

\end{abstract}

\end{titlepage}

\vskip 1cm

\section{Introduction}

The AdS/CFT correspondence is an invaluable tool \cite{Aharony:1999ti}, not
only enabling us to study the strong coupling limits of a wide variety of
gauge theories, but also allowing us to study deep questions arising in the
study of quantum gravity.

\vspace{1pt}Certain heavy supersymmetric states in the gauge theory can be
dual to geometric backgrounds in string theory. An example is the set of 1/2
BPS chiral primary states built with one complex scalar \cite{Corley:2001zk},%
\cite{Berenstein:2004kk},\cite{Lin:2004nb},\cite{Grant:2005qc},\cite%
{Horava:2005pv}. These operators can be labelled by Young tableaux or Schur
polynomials which themselves form an orthonormal basis. When the charges of
these operators are very large, they become dual to a variety of geometries,
which have bubbling type boundary conditions in some loci of the spacetime
far from the boundary, that encode the mapping from the gauge theory states.
Besides the 1/2 BPS chiral primary operators, there are also other types of
operators with corresponding bubbling geometries, e.g. \cite{Lunin:2006xr},\cite%
{Gomis:2007fi},\cite{Gomis:2006cu},\cite{Berenstein:2005aa} in $\mathcal{N}=4$ SYM.

A natural question which arises is what is the description of the non-BPS
excitations on top of the above states in both the gauge and string theory
sides? On the gauge side, we study operators which are products between a
Schur polynomial and a single trace operator. We may view the Schur
polynomial as the background geometry and the single trace part as a further
stringy excitation on the geometry. On the string side, we study closed
strings excited on the dual geometries, described by a general Schur
polynomial or Young tableau. In particular we study the backgrounds with
general concentric ring distributions \cite{Lin:2004nb} in the phase space
plane. The size of the $S^3$'s in the LLM metric are determined by the
position in the droplet distribution, in which the edges correspond to the
vanishing of one or the other $S^3$.

One way to study such non-BPS states is to look at the dilatation generator
acting on them. The dilatation operator acting on the non-BPS state is
naturally described in terms of the hamiltonian on a lattice, which itself
can be described in a variety of limits. The bosonic occupation numbers of
the lattice sites correspond to the number of complex scalar fields in the
trace part of the operator in between the impurity fields. This has been
studied before in the context of certain background geometries \cite%
{Vazquez:2006id},\cite{Chen:2007gh},\cite{Koch:2008ah},\cite{Koch:2008cm}.
We will be particularly interested in studying the coherent states of such a
lattice, though for more general background geometries. In the continuum
limit of the lattice we obtain a sigma model that matches with the sigma
model of the closed string living in the dual geometry. We can also study
the lattice with a small number of sites, corresponding to short BMN type
strings.

In \cite{Chen:2007gh} an analysis was made of a reduced subspace of possible
diagrams. In this work the authors analyzed the coherent state on a disk
plus ring distribution and proved that the normalized wavefunction
constructed from this coherent state knew about the geometry and topology of
the string theory background. In the current work we fully generalize this
description to include any geometries which can be described by axially
symmetric solutions in the phase space plane.

\vspace{1pt}Furthermore we will look at the entropy of particular ensembles
of such states. In the 1/2-BPS limit all non-equivalent geometries are
orthogonal to one another, meaning that an ensemble made of such solutions
does not mix. The breaking of supersymmetry via the stringy excitation/trace
operator allows for a mixing between elements of the gravitational ensemble
and a non-diagonal reduced density matrix.

The organization of this paper is as follows. In section 2.1, we introduce
the non-BPS operators that we will study, the lattice labelling, and the
associated oscillator algebra. In section 2.2, we analyze the coherent state
wavefunction of the system, the continuum limit of the lattice and sigma
models, and also the properties of the coherent state wavefunction, such as
the norm and the average occupation number. In section 2.3, we describe a
density matrix over these coherent states with a constrained set of Young
tableaux, and from this we are able to calculate the entropy of such an
ensemble of states. In section 2.4, we derive the BMN states from the ansatz
of the coherent states and show that in the appropriate limits we recover
the results of \cite{Chen:2007gh}. Finally, in section 3, we briefly discuss
our results and conclude. We also include four appendices providing concrete
computational details on the calculations discussed in the body of the text.

\section{Strings on Bubbling Geometries}

\label{strings_on_bubbling_geometries}

\renewcommand{\theequation}{2.\arabic{equation}} \setcounter{equation}{0}

\subsection{Operators}

\label{operators}

The 1/2 BPS chiral primary operators of $\mathcal{N}=4$ super Yang Mills can
be labelled by Schur Polynomials $\chi _{R}(\mathcal{Z})$, where $R$ is the
representation in which the operator is being traced, and $\mathcal{Z}$ is one of the complex
scalar fields. Equivalently they can be represented by Young tableaux which
are dual to $1/2$ BPS geometries. These can be described using the LLM
prescription as a series of concentric black and white annuli distributions.
The Young tableau corresponding to a particular chiral primary operator may
be labelled by the positions of its corners. We label outward corners with a
``$b$" and inward corners with an ``$a$", going from bottom left to top
right. A given diagram will come with labels $%
b_{1},a_{1},b_{2},a_{2},...,b_{m},a_{m}$ where $m$ is the total number of
inward, or equivalently outward pointing corners. Such operators can be
mapped into a distribution of $m$ black rings where the inner and outer
edges of a ring/annulus correspond on an outward pointing or inward pointing
edge of a Young tableau, respectively. We label these inner and outer radii
as $R_{2k-1}$ and $R_{2k}$ where $k$ labels the ring (we will use $k$ and $l$
to label ring numbers throughout). Due to quantization of the phase space
area, the radius squared $R_{2k-1}^{2}~$of each inner circle of the black
rings is proportional to the distance $C_{b_{k}}$ from the lower-left point
of the Young tableau to the outward corner $b_{k}$, of the Young tableau,
counting along the edge. Similarly, the radius squared $R_{2k}^{2}~$of each
outer circle of the black rings is proportional to the distance $C_{a_{k}}$
from the lower-left point of the Young tableau to the inward corner $a_{k}$ 
\footnote{%
The coordinates of the $z$-plane in this paper are rescaled from the
coordinates of the $z$-plane in \cite{Lin:2004nb} by a factor, such that $%
\frac{R^{2}|_{here}}{1/N}=\frac{R^{2}|_{there}}{2\hbar }.$}:%
\begin{eqnarray}
R_{2k-1} &=&\frac{\sqrt{C_{b_{k}}}}{\sqrt{N}},~~~~R_{2k}=\frac{\sqrt{%
C_{a_{k}}}}{\sqrt{N}}\,,  \label{radius_distance_relation} \\
C_{b_{k}} &=&C_{a_{k}}-N_{k},~\ \ \
C_{a_{k}}=\sum\limits_{j=1}^{k}(M_{j}+N_{j})\,.
\end{eqnarray}%
This provides a simple mapping between a Young tableau labelling and a phase
space distribution. 
\begin{figure}[h]
\begin{center}
\includegraphics[width=14cm]{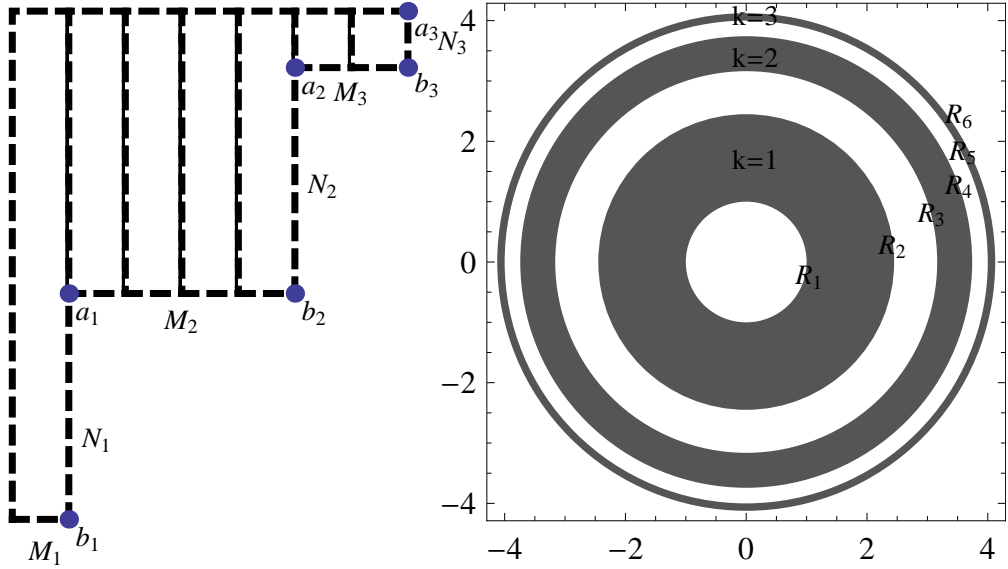}
\end{center}
\caption{{\protect\small An example of the Young tableau description of a
Schur polynomial operator in the gauge theory (with small $N$ and $M$ as an
example). The numbers of rows and columns on each step ($M_{k}$ and $N_{k}$)
are labelled along with the positions of the inner and outer edges of the
associated annulus distribution in the LLM plane. The rings are labelled $%
k=1 $ to $3$ from the inner ring to the outer ring. The dark regions in the
LLM plane are those with boundary value $-\frac{1}{2}$ but are shaded in
grey for clarity. The ring radii squared are: $(1,5,9,13,15,16)$.}}
\label{fig.Youngrings}
\end{figure}

It should be noted that such Young tableaux do not exhaust all possible
concentric ring distributions. Such a description will not include phase
space distributions which start with a black disk\footnote{%
These would correspond to ${M_{1}}=0,{R_{1}}=0~$limit, which will be
discussed more in Appendix \ref{appendix_coherent_state_derivation}.}. These
are labelled separately with a thin vertical line running from the bottom
left hand corner of the Young tableau, corresponding to having no white
region in the centre. The labelling of such states follows logically. The
bulk of this paper is motivated by studying closed string states in backgrounds given by general
axially symmetric ring distributions.

Having described this subspace of the half-BPS sector we would like to look
at products of a trace multiplied by an operator
described by the Schur polynomials discussed above. Such products are in
general non-BPS and we will utilize the breaking of supersymmetry in what
follows. We label such products as: 
\begin{equation}
\mathcal{O}_{\{n_{j}\}}=tr\left( \prod\limits_{j=1}^{L}\mathcal{W}_{\alpha
_{j}}\mathcal{Z}_{a_{j}}^{n_{j}}\right) \chi _{R}(\mathcal{Z})\,,
\label{operator_label}
\end{equation}%
where $\chi _{R}(\mathcal{Z})$ is a Schur polynomial corresponding to a
Young tableau of representation $R$. \ $\mathcal{W}_{\alpha _{j}}$ is a
letter corresponding to impurities coming from fields which are not in the
1/2 BPS chiral primary sector. $\mathcal{Z}_{a_{j}}$ is a field creating or removing
a box with weight $C_{a_k}$ or $C_{b_k}$ at the Young tableau corner, where the subscript $a_j$ denotes the corner $a_k$ or $b_k$ (see also \cite{Koch:2008ah}). Such an operator corresponds to a closed string, represented by the trace, in a background represented by the Schur
polynomial. The presence of the string in the background leads to SUSY
breaking. The operators that are eigenstates of the dilatation operator are
superpositions in the space spanned by the basis (\ref{operator_label}). (In \cite{Chen:2007gh}, product of determinants were used in place of the Schur polynomial part.) The
total number of $\mathcal{Z}$s in the trace part of (\ref{operator_label})
is not fixed when we mix these operators to obtain the eigenstates of the
dilatation generator, so it is convenient to view the trace part of (\ref%
{operator_label}) as a lattice of $L$ sites each with an occupation number $%
n_{j}$. For simplicity, we first take $\mathcal{W}_{\alpha _{j}}$ to come
purely from the complex scalar $Y$. Each operator can thus be represented by
a state $\left\vert n_{1},n_{2},...,n_{L}\right\rangle .$ The positive
occupation number corresponds to positive angular momentum of the string
along the outer circle of a black ring, while the negative occupation number
corresponds to the negative momentum of the string along the inner circle of
the black ring. Such a basis can be used to describe a wide variety of
string states in 1/2-BPS backgrounds.

The effective Hamiltonian of the lattice is derived from the dilatation
generator $\Delta =\Delta _{0}+\lambda \Delta _{1}+\sum\limits_{n\geq
2}\lambda ^{n}{\hat{\Delta}}_{n}$, where

\begin{equation}
\Delta _{0}+\lambda \Delta _{1}=tr(Z\partial _{Z}+Y\partial _{Y})+\frac{%
g_{YM}^{2}}{8\pi }tr[Z,Y][\partial _{Z},\partial _{Y}]\,,
\end{equation}%
for the operators under consideration, and here $\lambda =\frac{g_{YM}^{2}N}{%
8\pi }.$ We have%
\begin{eqnarray}
\ \Delta _{0} &=&(J_{\chi }+\sum\limits_{j=1}^{L}n_{j})+L=\mathcal{J}+L\,, \\
\lambda D_{1} &=&\frac{g_{YM}^{2}}{8\pi }tr[Z,Y][\partial _{Z},\partial
_{Y}]=\Delta -\mathcal{J}-L=H\,,
\end{eqnarray}%
where in the last line we only consider the one-loop dilation generator. The
Hamiltonian when acting on the lattice takes the form%
\begin{equation}
H=E-\mathcal{J}-L=\lambda ~\sum\limits_{j=1}^{L}(a_{j}^{\dagger
}-a_{j+1}^{\dagger })(a_{j}-a_{j+1})\,,  \label{hamiltonian_label}
\end{equation}%
where $a_{j}$ are shift operators which reduce the bosonic occupation number
by one. We use $\mathcal{J}$ to denote the total $U(1)$-charge associated
with $\mathcal{Z,}$ and use $J$ to denote that $U(1)$-charge from the trace
part, i.e. 
\begin{equation}
J=J_{trace}=\sum\limits_{j=1}^{L}n_{j}.
\end{equation}

For a general Young Tableau, we have the following algebra%
\begin{eqnarray}
a\left\vert -n,b_{k}\right\rangle &=&\frac{\sqrt{C_{b_{k}}}}{\sqrt{N}}%
\left\vert -n-1,b_{k}\right\rangle ,~n\geqslant 0,~ \\
a\left\vert 1,a_{l}\right\rangle &=&\overset{m}{\underset{k=1}{\sum }}%
v_{l}^{k}\frac{\sqrt{C_{a_{k}}}}{\sqrt{N}}\left\vert 0\right\rangle _{k},~~\
a\left\vert n,a_{k}\right\rangle =\frac{\sqrt{C_{a_{k}}}}{\sqrt{N}}%
\left\vert n-1,a_{k}\right\rangle ,~~n>1 \\
a^{\dagger }\left\vert n,a_{k}\right\rangle &=&\frac{\sqrt{C_{a_{k}}}}{\sqrt{%
N}}\left\vert n+1,a_{k}\right\rangle ,~~~n\geqslant 0,~~~~\ \ \left\vert
0,b_{k}\right\rangle =\left\vert 0,a_{k}\right\rangle =\left\vert
0\right\rangle _{k}\,, \\
~a^{\dagger }\left\vert -n,b_{k}\right\rangle &=&\frac{\sqrt{C_{b_{k}}}}{%
\sqrt{N}}\left\vert -n+1,b_{k}\right\rangle ,~n>0.
\end{eqnarray}%
The state $\left\vert -n,b_{k}\right\rangle $ corresponds to the $-n^{th}$
excitation on the inner edge of the $k^{th}$ ring while $\left\vert
n,a_{k}\right\rangle $ corresponds to the $n^{th}$ excitation on the outer
edge of the $k^{th}$ ring. As noted before, the $C_{b_{k}}$ and $C_{a_{k}}$
are the distances (number of boxes) between the $k^{th}$ outward pointing or
inward pointing corner and the bottom-left point of the Young tableau. The $%
C_{b_{k}}$ and $C_{a_{k}}~$are also the weights of the boxes of these Young
tableau corners, and are the coefficients that appear in the two-point
correlators for the fields $\mathcal{Z}_{a_{j}}$ \cite{Koch:2008ah}. The
action of the annihilation/creation operator on different types of $\mathcal{%
Z}_{a_{j}}$ field excitations differ by the coefficients $\sqrt{C_{b_{k}}/N}$%
, or $\sqrt{C_{a_{k}}/N}$, and is otherwise similar. The action of $a_{j}$ ($%
a_{j}^{\dagger }$) removes (adds) a box in the appropriate corner of the
Young tableau. More discussions of their properties are included in Appendix %
\ref{appendix_coherent_state_derivation}.

In principle one can take several limits in order to study superpositions of
operators of the form (\ref{operator_label}), and their string theory duals.
The limits of interest (both on the gauge and string side) include:

1. The continuous limit/sigma model limit:%
\begin{equation}
L\gg 1,~~\frac{\lambda }{L^{2}}\ll 1~,~~J\;\mathrm{unfixed};~\ ~~\lambda \ll
1,~~~\text{v.s.}~~~\frac{1}{\lambda }\ll 1
\end{equation}

2. The BMN/plane-wave type limit:%
\begin{equation}
L=2,~\ \ ~~|J|\gg 1,~~\frac{\lambda }{J^{2}}\ll 1~;~\ ~~\lambda \ll 1,\quad 
\text{v.s.}~~~~\frac{1}{\lambda }\ll 1
\end{equation}

3. The Hofman-Maldacena type limit:%
\begin{equation}
L=2,~~~~~|J|\gg 1,~~\frac{n}{|J|}~\mathrm{finite};~\ ~~\lambda ~\mathrm{%
finite}
\end{equation}%
where in the last two cases, we use a \vspace{1pt}two-site lattice,
corresponding to a two-magnon state. It is of course possible to study this
in a more general $L$-site lattice.

\subsection{Coherent String States}

\label{coherent_string_states}

We would like to be able to describe a coherent
string state on the background given by an arbitrary distribution of black
rings, as discussed in the previous section. In order to be able to do this
we find a coherent string prescription which is normalizable on each of the $%
m$ black rings separately. The coherent state satisfies the equation \cite%
{Zhang:1990fy} 
\begin{equation}
a\left\vert z\right\rangle _{l}=z\left\vert z\right\rangle _{l}\,,
\end{equation}%
where the $l$ labels the $l^{th}$ ring in which the wavefunction is well
defined and normalizable.

The coherent state can be expressed as%
\begin{equation}
\left\vert z\right\rangle _{l}=\overset{l}{\underset{k=1}{\sum }}f_{l}^{2k-1}%
\overset{\infty }{\underset{n=0}{\sum }}\left( \frac{R_{2k-1}}{z}\right)
^{n}\left\vert -n,b_{k}\right\rangle +\overset{m}{\underset{k=l}{\sum }}%
f_{l}^{2k}\overset{\infty }{\underset{n=1}{\sum }}\left( \frac{z}{R_{2k}}%
\right) ^{n}\left\vert n,a_{k}\right\rangle \,,  \label{coherent_state_l}
\end{equation}%
which is normalizible on the $l^{th}$ black ring, i.e. $R_{2l-1}<|z|<R_{2l}$%
. The state $\left\vert n,a_{k}\right\rangle $ corresponds to the $n^{th}$
excitation on the outer edge of the $k^{th}$ ring while $\left\vert
-n,b_{k}\right\rangle $ corresponds to the $-n^{th}$ excitation on the inner
edge of the $k^{th}$ ring. It can be shown that the sums over $k$ can be
consistently restricted to the ranges in (\ref{coherent_state_l}) due to the
normalizability on the $l^{th}$ ring. A more detailed derivation of the
above is included in Appendix \ref{appendix_coherent_state_derivation}.

In (\ref{coherent_state_l}), we sum over all the positive occupancies, or in
other words, positive angular momentum movers along the outer circles,
outside the $l^{th}$ black ring, and over all the negative occupancies, or,
negative angular momentum movers along the inner circles, inside the $l^{th}$
black ring. Note that such a summation is important for normalizability
within the $l^{th}$ ring. The coefficients $f_{l}^{2k-1}$ and $f_{l}^{2k}$
correspond to the relative contribution to the wavefunction from the
excitations associated with the $(2k-1)^{th}$ and $2k^{th}~$circles.
Equivalently these circles correspond to the $b_{k},~a_{k}~$corners of the
Young tableau from the gauge theory point of view.

(\ref{coherent_state_l}) is a Fock space state which sums over all
occupation numbers, as well as all types of oscillators specified by the
Young tableaux corners $b_{k},a_{k}$. The real, positive coefficients $%
f_{l}^{2k-1},f_{l}^{2k}$ can be shown to satisfy the following relation 
\begin{equation}
\overset{m}{\underset{k=l}{\sum }}f_{l}^{2k}v_{k}^{j}=f_{l}^{2j-1},~~j%
\leqslant l;~~~\ \ \ \ \overset{m}{\underset{k=l}{\sum }}%
f_{l}^{2k}v_{k}^{j}=0,~~j>l \, ,
\end{equation}%
where $v_k^j$ is a mixing matrix which allows for mixing between different
rings. The $f^{\prime}s$ can then be written as

\begin{eqnarray}
(f_{l}^{2k-1})^{2} &=&\frac{-c_{l}%
\prod_{p=l+1}^{m}(R_{2k-1}^{2}-R_{2p-1}^{2})}{\prod_{p=1\neq
k}^{l}(R_{2k-1}^{2}-R_{2p-1}^{2})}\frac{%
\prod_{p=1}^{l-1}(R_{2k-1}^{2}-R_{2p}^{2})}{%
\prod_{p=l}^{m}(R_{2k-1}^{2}-R_{2p}^{2})} \\
&=&\frac{-c_{l}N\prod_{p=l+1}^{m}(C_{b_{k}}-C_{b_{p}})}{\prod_{p=1\neq
k}^{l}(C_{b_{k}}-C_{b_{p}})}\frac{\prod_{p=1}^{l-1}(C_{b_{k}}-C_{a_{p}})}{%
\prod_{p=l}^{m}(C_{b_{k}}-C_{a_{p}})}\,,
\end{eqnarray}%
\begin{eqnarray}
(f_{l}^{2k})^{2} &=&\frac{c_{l}\prod_{p=l+1}^{m}(R_{2k}^{2}-R_{2p-1}^{2})}{%
\prod_{p=1}^{l}(R_{2k}^{2}-R_{2p-1}^{2})}\frac{\prod_{p=1\neq
k}^{l-1}(R_{2k}^{2}-R_{2p}^{2})}{\prod_{p=l\neq k}^{m}(R_{2k}^{2}-R_{2p}^{2})%
} \\
&=&\frac{c_{l}N\prod_{p=l+1}^{m}(C_{a_{k}}-C_{b_{p}})}{%
\prod_{p=1}^{l}(C_{a_{k}}-C_{b_{p}})}\frac{\prod_{p=1\neq
k}^{l-1}(C_{a_{k}}-C_{a_{p}})}{\prod_{p=l\neq k}^{m}(C_{a_{k}}-C_{a_{p}})}\,,
\label{eq.fs}
\end{eqnarray}%
where we have used the identity (\ref{radius_distance_relation}) to write
these expressions in terms of $C_{b_{k}},C_{a_{k}}$. It can be shown that
these expressions are positive definite. \vspace{1pt} By looking at the
behavior of the norm $\left\langle z|z\right\rangle $ when merging two black
rings, we get%
\begin{equation}
c_{l}=\frac{c_{R}\overset{l}{\underset{k=2}{\prod }}R_{2k-1}^{2}\overset{m}{%
\underset{k=l}{\prod }}R_{2k}^{2}}{\overset{l-1}{\underset{k=1}{\prod }}%
R_{2k}^{2}\overset{m}{\underset{k=l+1}{\prod }}R_{2k-1}^{2}}=\frac{c_{R}%
\overset{l}{\underset{k=2}{\prod }}C_{b_{k}}\overset{m}{\underset{k=l}{\prod 
}}C_{a_{k}}}{N\overset{l-1}{\underset{k=1}{\prod }}C_{a_{k}}\overset{m}{%
\underset{k=l+1}{\prod }}C_{b_{k}}}\,,  \label{eq.cl}
\end{equation}%
\vspace{1pt}where $c_{R}$ is an overall normalization which is unfixed from
the coherent state condition but will be fixed later by putting a constraint
on the integral over the inner product of such representations in the $z$%
-plane. Finally, one can write the coherent state wave-function on the
entire $z$-plane as a sum over the states defined on each ring: \vspace{1pt}%
\begin{equation}
\left\vert z\right\rangle =\overset{m}{\underset{l=1}{\sum }}\left\vert
z\right\rangle _{l}~\theta _{R_{2l-1},R_{2l}}\,,  \label{eq.z}
\end{equation}%
where $\theta _{x,y}$ has unit support between $x$ and $y$, and is 0
otherwise. \vspace{1pt}

Having defined such a coherent state on the 1/2 BPS background, one can take
the continuum limit: 
\begin{equation}
L\gg 1,~~\frac{\lambda }{L^{2}}\ll 1~;~\ ~~\lambda \ll 1\,,
\end{equation}%
as the product of coherent states on each lattice site: 
\begin{equation}
a_{j}\left\vert z_{j}\right\rangle =z_{j}\left\vert z_{j}\right\rangle \,.
\end{equation}%
This limit would correspond to the large angular momentum ($L$) limit on the string side. A
general coherent state is then labelled by a coherent state on each site of
the lattice: 
\begin{equation}
\left\vert z_{1},z_{2},...,z_{L}\right\rangle \,.
\end{equation}%
\vspace{1pt}The continuum limit can be expressed in terms of a sigma model
action with Hamiltonian (\ref{hamiltonian_label}), which is derived by
looking at the large $L$ limit of the lattice action:%
\begin{eqnarray}
S &=&\int dt\left( \left\langle z_{1},z_{2},...,z_{L}\right\vert i\frac{d}{dt%
}\left\vert z_{1},z_{2},...,z_{L}\right\rangle -\left\langle
z_{1},z_{2},...,z_{L}\right\vert H\left\vert
z_{1},z_{2},...,z_{L}\right\rangle \right)  \notag \\
&=&\int dt\sum\limits_{j=1}^{L}(\left\langle z_{j}\right\vert i\frac{d}{dt}%
\left\vert z_{j}\right\rangle -\lambda \left\vert z_{j}-z_{j+1}\right\vert
^{2})  \notag \\
&=&L\int dt\int_{0}^{1}d\sigma \left( \frac{i}{2}(\overset{.}{\overline{z}}%
\partial _{\overline{z}}-\overset{.}{z}\partial _{z})\log (\left\langle
z|z\right\rangle )-\frac{\lambda }{L^{2}}|z^{\prime }|^{2}\right) \,, \\
V(z,\overline{z}) &=&\partial _{\overline{z}}\log (\left\langle
z|z\right\rangle )\,.
\end{eqnarray}%
This is a sigma model with target space the $z$-plane of the 1/2 BPS
bubbling chiral primary geometry \cite{Vazquez:2006id},\cite{Chen:2007gh}, see as well \cite%
{Koch:2008ah}. This Landau-Lifshitz type action coincides with the action
from the string theory side for the string sigma model that describes a
closed string traveling with large angular momentum $L$ on the $S^{3}$ in
the 1/2 BPS bubbling geometry under the uniform gauge, in which the angular
momenta on the $S^{3}$ are uniformly distributed along the string. This
uniform gauge corresponds to the uniform distribution of the $Y$ fields on
the lattice sites. This action can be written in terms of the above
potential as: 
\begin{eqnarray}
S &=&\frac{L}{\alpha ^{\prime }}\int dt\int_{0}^{1}d\sigma \left( \frac{i}{2}%
(V\overset{.}{\overline{z}}-\overset{.}{z}\overline{V})-\frac{\lambda }{L^{2}%
}|z^{\prime }|^{2}\right)  \label{sigma_model_string_side} \\
&=&\frac{L}{\alpha ^{\prime }}\int dt\int_{0}^{1}d\sigma \left( h^{2}\frac{i%
}{2}(z\overset{.}{\overline{z}}-\overset{.}{z}\overline{z})-\frac{\lambda }{%
L^{2}}|z^{\prime }|^{2}\right) \\
&=&\frac{L}{\alpha ^{\prime }}\int dt\int_{0}^{1}d\sigma \left( h^{2}\rho
^{2}\overset{.}{\phi }-\frac{\lambda }{L^{2}}(\rho ^{\prime 2}+\rho ^{2}\phi
^{\prime 2})\right) \,,
\end{eqnarray}%
\begin{equation}
V=h^{2}z,~~~~\ ~z=\rho e^{i\phi }\,.
\end{equation}%
We have neglected higher order terms $\left( \frac{\lambda }{L^{2}}\right)
^{2}|z^{\prime }|^{4},~\left( \frac{\lambda }{L^{2}}\right) ^{2}\frac{%
|z^{\prime \prime }|^{2}}{h^{4}}$, which would correspond to the
contribution from the two-loop dilatation operators and can be safely
neglected in this limit.

The equations of motion coming from this action are%
\begin{equation}
\frac{\lambda }{L^{2}}z^{\prime \prime }=\frac{i}{2}(\partial _{\overline{z}}%
\overline{V}+\partial _{z}V)\overset{.}{z}\, ,
\end{equation}%
which have static solution%
\begin{eqnarray}
\overset{.}{\phi } &=&0,~\ \overset{.}{\rho }=0,~\ \ \phi ^{\prime
}=0,~~\rho ^{\prime }=c_{1}\, ,~ \\
~~z &=&(c_{1}\sigma +c_{3})e^{ic_{2}}\, ,
\end{eqnarray}%
where $c_{1},c_{2},c_{3}$ are constants of motion, and the solution
corresponds to a straight string stretching across a black ring (and back
again).

One can also bring (\ref{sigma_model_string_side}) into the
canonical form of a 2d complex scalar field theory. We can transform the $%
z\rightarrow \varphi ,$ bringing the kinetic term into its canonical form%
\vspace{1pt}%
\begin{eqnarray}
&&h^{2}\frac{i}{2}(z\overset{.}{\overline{z}}-\overset{.}{z}\overline{z}) =%
\frac{i}{2}(\varphi \overset{.}{\overline{\varphi }}-\overset{.}{\varphi }%
\overline{\varphi }), \\
&& z =\rho e^{i\phi },~~\ ~~\varphi =re^{i\phi }, \\
&& h\rho =r,~~~\ \ \ \ ~\varphi =hz\,.
\end{eqnarray}%
Under this transformation, the complex scalar field theory is described by
the action \vspace{1pt} 
\begin{equation}
S=\frac{L}{\alpha ^{\prime }}\int dtd\sigma \left[ \frac{i}{2}(\varphi 
\overset{.}{\overline{\varphi }}-\overset{.}{\varphi }\overline{\varphi })-%
\frac{\lambda }{L^{2}}\left( f(|\varphi |^{2})\frac{1}{2}|\varphi ^{\prime
}|^{2}+g(|\varphi |^{2})\frac{(\varphi \overline{\varphi }^{\prime })^{2}+(%
\overline{\varphi }\varphi ^{\prime })^{2}}{4|\varphi |^{2}}\right) \right]
\,,
\end{equation}%
where 
\begin{equation}
f(|\varphi |^{2})=(\partial _{r}\rho )^{2}+\frac{1}{h^{2}},~~~\ ~g(|\varphi
|^{2})=(\partial _{r}\rho )^{2}-\frac{1}{h^{2}}\,.
\end{equation}%
The potential terms in this action are dependent on the background in which
the string is moving. This means that the action is dependent directly on
the droplet configuration in the $z$-plane. In particular the topology of
the background which is related to the number of inner and outer rings is
captured in the potential. One can study the S-matrix of this model, via
similar methods to \cite{Roiban:2006yc} for $AdS_{5}\times S^{5}$. For
example for the $AdS_{5}\times S^{5}~$case the potential is given by:

\vspace{1pt} \vspace{1pt} 
\begin{equation}
\frac{V}{z}=\frac{1}{|R_{1}^{2}-|z|^{2}|}\, ,
\end{equation}

\begin{equation}
z=\frac{R_{1}r}{\sqrt{1+r^{2}}}e^{i\phi },~~~\ \ ~\varphi =re^{i\phi }\, ,
\end{equation}%
$z\rightarrow \varphi ~$maps the disk into a whole plane and $R_1$ is the
radius of the disk. For more general concentric black-ring configurations we have

\begin{equation}
\frac{V}{z}=\overset{m}{\underset{k=1}{\sum }}\left[ -\frac{1}{\left\vert
R_{2k-1}^{2}-\left\vert z\right\vert ^{2}\right\vert }+\frac{1}{\left\vert
R_{2k}^{2}-\left\vert z\right\vert ^{2}\right\vert }\right]\, .
\end{equation}%
The limits of such an expression can be found in Appendix \ref%
{appendix_coherent_state_derivation}.

\vspace{1pt} We would now like to derive such a potential from the coherent
state. We first consider the norm of the coherent states $\left\langle
z|z\right\rangle .~$ The coherent state (\ref{coherent_state_l}) defined on
the $l^{th}$ ring is: 
\begin{equation}
\left\vert z\right\rangle _{l}=\overset{l}{\underset{k=1}{\sum }}%
f_{l}^{2k-1}\left\vert z,b_{k}\right\rangle +\overset{m}{\underset{k=l}{\sum 
}}f_{l}^{2k}\left\vert z,a_{k}\right\rangle \,,
\end{equation}%
where 
\begin{equation}
\left\vert z,b_{k}\right\rangle =\overset{\infty }{\underset{n=0}{\sum }}%
\left( \frac{R_{2k-1}}{z}\right) ^{n}\left\vert -n,b_{k}\right\rangle
,~~~~~~~\left\vert z,a_{k}\right\rangle =\overset{\infty }{\underset{n=1}{%
\sum }}\left( \frac{z}{R_{2k}}\right) ^{n}\left\vert n,a_{k}\right\rangle \,,
\end{equation}%
which have the property%
\begin{eqnarray}
\left\langle z,b_{k}|z,b_{k}\right\rangle &=&-\frac{R_{2k-1}^{2}}{%
R_{2k-1}^{2}-\left\vert z\right\vert ^{2}}+1=-\frac{\left\vert z\right\vert
^{2}}{R_{2k-1}^{2}-\left\vert z\right\vert ^{2}}\,, \\
\left\langle z,a_{k}|z,a_{k}\right\rangle &=&\frac{R_{2k}^{2}}{%
R_{2k}^{2}-\left\vert z\right\vert ^{2}}-1=\frac{\left\vert z\right\vert ^{2}%
}{R_{2k}^{2}-\left\vert z\right\vert ^{2}}\,.
\end{eqnarray}%
\vspace{1pt}\vspace{1pt} From the coherent state expansion we get%
\begin{equation}
\left\langle z|z\right\rangle _{l}=\overset{l}{\underset{k=1}{\sum }}-\frac{%
(f_{l}^{2k-1})^{2}R_{2k-1}^{2}}{R_{2k-1}^{2}-\left\vert z\right\vert ^{2}}+%
\overset{m}{\underset{k=l}{\sum }}\frac{(f_{l}^{2k})^{2}R_{2k}^{2}}{%
R_{2k}^{2}-\left\vert z\right\vert ^{2}}\,~,  \label{coherent_state_norm_1}
\end{equation}%
which leads to a relation between the coefficients in the coherent state
expansion 
\begin{equation}
\overset{l}{\underset{k=1}{\sum }}-(f_{l}^{2k-1})^{2}+\overset{m}{\underset{%
k=l}{\sum }}(f_{l}^{2k})^{2}=0\,,
\end{equation}%
Note that the coherent state norm is nonzero only on the black rings. The
expression (\ref{coherent_state_norm_1}) can be summed and factorized in the
numerator, and rewritten as 
\begin{eqnarray}
\left\langle z|z\right\rangle _{l} &=&\frac{c_{l}\overset{l-1}{\underset{k=1}%
{\prod }}\left( R_{2k}^{2}-\left\vert z\right\vert ^{2}\right) \overset{m}{%
\underset{k=l+1}{\prod }}\left( R_{2k-1}^{2}-\left\vert z\right\vert
^{2}\right) (-\left\vert z\right\vert ^{2})}{\overset{l}{\underset{k=1}{%
\prod }}\left( R_{2k-1}^{2}-\left\vert z\right\vert ^{2}\right) \overset{m}{%
\underset{k=l}{\prod }}\left( R_{2k}^{2}-\left\vert z\right\vert ^{2}\right) 
}  \label{coheren_state_norm_2} \\
&=&\frac{c_{R}\overset{l-1}{\underset{k=1}{\prod }}\left( 1-\frac{\left\vert
z\right\vert ^{2}}{R_{2k}^{2}}\right) \overset{m}{\underset{k=l+1}{\prod }}%
\left( 1-\frac{\left\vert z\right\vert ^{2}}{R_{2k-1}^{2}}\right) }{\overset{%
l}{\underset{k=2}{\prod }}\left( 1-\frac{\left\vert z\right\vert ^{2}}{%
R_{2k-1}^{2}}\right) \overset{m}{\underset{k=l}{\prod }}\left( 1-\frac{%
\left\vert z\right\vert ^{2}}{R_{2k}^{2}}\right) }\frac{(-\left\vert
z\right\vert ^{2})}{\left( R_{1}^{2}-\left\vert z\right\vert ^{2}\right) }\,,
\end{eqnarray}%
The norm of the coherent state can also be written in terms of the
potential, $V$, as%
\begin{equation}
\left\langle z|z\right\rangle _{l}=e^{\int V(z,\overline{z})d\overline{z}}\,,
\label{coherent_state_norm_integral}
\end{equation}%
which can be used to calculate the potential: 
\begin{eqnarray}
V(z,\overline{z}) &=&\overset{l-1}{\underset{k=1}{\sum }}\left[ \frac{z}{%
R_{2k-1}^{2}-\left\vert z\right\vert ^{2}}-\frac{z}{R_{2k}^{2}-\left\vert
z\right\vert ^{2}}\right] +\overset{m}{\underset{k=l+1}{\sum }}\left[ -\frac{%
z}{R_{2k-1}^{2}-\left\vert z\right\vert ^{2}}+\frac{z}{R_{2k}^{2}-\left\vert
z\right\vert ^{2}}\right]  \notag \\
&&+\frac{z}{R_{2l-1}^{2}-\left\vert z\right\vert ^{2}}+\frac{z}{%
R_{2l}^{2}-\left\vert z\right\vert ^{2}}\,.
\end{eqnarray}%
Note that the existence of $c_{R}$ is consistent with the existence of an
integration constant in (\ref{coherent_state_norm_integral}).

The coherent state norm is non-zero only on the black-rings as
the potential diverges at their edges. This calculation of the potential
matches with the $V$ function derived in the sigma model from the string
theory side. This shows that the coherent state wavefunction knows about the
geometry of the multi-ring background. Moreover, the coherent state
wavefunction has a different expression on each disconnected black ring,
meaning that it knows about the disconnectedness of the droplets, $\mathit{i.e.}$ topology,
of the background.

Another property we can study using the coherent state we have described
above is the average value of the occupation number on each lattice site.
From (\ref{coherent_state_l}), we have the coefficients in the coherent
state expansion expressed as 
\begin{equation}
f_{l}^{2k-1}=\frac{\left\langle z,b_{k}|z\right\rangle _{l}}{\left\langle
z,b_{k}|z,b_{k}\right\rangle },~~f_{l}^{2k}=\frac{\left\langle
z,a_{k}|z\right\rangle _{l}}{\left\langle z,a_{k}|z,a_{k}\right\rangle },~~%
\frac{R_{2k-1}^{2n}}{\left\vert z\right\vert ^{2n}}=\left\langle
z,b_{k}|-n,b_{k}\right\rangle ^{2},~~\frac{\left\vert z\right\vert ^{2n}}{%
R_{2k}^{2n}}=\left\langle z,a_{k}|n,a_{k}\right\rangle ^{2}\,.
\end{equation}%
The states are orthonormal, such that%
\begin{equation}
\left\langle -n_{1},b_{k1}|-n_{2},b_{k2}\right\rangle =1\cdot \delta
_{n_{1},n_{2}}\delta _{b_{k1},b_{k2}},~~~\left\langle
n_{1},a_{k1}|n_{2},a_{k2}\right\rangle =1\cdot \delta _{n_{1},n_{2}}\delta
_{a_{k1},a_{k2}}\,.
\end{equation}%
Excitations of a different level or on a different ring are orthogonal.

The coherent state is a superposition of states with different occupation
numbers so in order to find the average occupancy we have to look at the
probability of occupancy $-n$ or $n$ on a given inner or outer ring. For the
inner rings the probability $P(-n,b_{k})$ of occupation number $-n$ on the $%
k^{th}$ ring is given by: 
\begin{equation}
P(-n,b_{k}) =\frac{\left\langle z,b_{k}|-n,b_{k}\right\rangle ^{2}}{%
\left\langle z,b_{k}|z,b_{k}\right\rangle }=\frac{R_{2k-1}^{2n}}{\left\vert
z\right\vert ^{2n}}\left( 1-\frac{R_{2k-1}^{2}}{\left\vert z\right\vert ^{2}}%
\right)\, ,
\end{equation}%
and the average occupation number $\bar{n}_{b_{k}}$ on the inner edge of the 
$k^{th}$ ring is (for $n\geqslant 0$)%
\begin{equation}
\bar{n}_{b_{k}} =\overset{\infty }{\underset{n=0}{\sum }}(-n)P(-n,b_{k})=-%
\frac{1}{\frac{\left\vert z\right\vert ^{2}}{R_{2k-1}^{2}}-1}\, .
\end{equation}

Similarly for the outer edge of the $k^{th}$ ring we have a probability $%
P(n,a_{k})$ of occupation number $n$: 
\begin{equation}
P(n,a_{k})=\frac{\left\langle z,a_{k}|n,a\right\rangle ^{2}}{\left\langle
z,a_{k}|z,a_{k}\right\rangle }=\frac{\left\vert z\right\vert ^{2(n-1)}}{%
R_{2k}^{2(n-1)}}\left( 1-\frac{\left\vert z\right\vert ^{2}}{R_{2k}^{2}}%
\right) \,,
\end{equation}%
and the average occupation number $\bar{n}_{a_{k}}$ on the outer edge of the 
$k^{th}$ ring is (for $n>0$)%
\begin{equation}
\bar{n}_{a_{k}}=\overset{\infty }{\underset{n=1}{\sum }}n\cdot P(n,a_{k})=%
\frac{1}{1-\frac{\left\vert z\right\vert ^{2}}{R_{2k}^{2}}}\,.
\end{equation}

We can now define an occupation number operator $\hat{n}$. The average
occupation over all inner and outer rings (corresponding to the average $%
U(1) $ $J$ charge on the string) for the entire coherent state is 
\begin{align}
\left\langle n\right\rangle _{l}& =\frac{\left\langle z\mid \hat{n}\mid
z\right\rangle _{l}}{\left\langle z|z\right\rangle _{l}}=\frac{\overset{l}{%
\underset{k=1}{\sum }}\,\overset{\infty }{\underset{n=0}{\sum }}\left\langle
-n,b_{k}|z\right\rangle _{l}^{2}(-n)+\overset{m}{\underset{k=l}{\sum }}\,%
\overset{\infty }{\underset{n=1}{\sum }}\left\langle n,a_{k}|z\right\rangle
_{l}^{2}(n)}{\left\langle z|z\right\rangle _{l}}
\label{average_occupation_1} \\
& =\frac{\overset{l}{\underset{k=1}{\sum }}(f_{l}^{2k-1})^{2}\left\langle
z,b_{k}|z,b_{k}\right\rangle \bar{n}_{b_{k}}+\overset{m}{\underset{k=l}{\sum 
}}(f_{l}^{2k})^{2}\left\langle z,a_{k}|z,a_{k}\right\rangle \bar{n}_{a_{k}}}{%
\left\langle z|z\right\rangle _{l}} \\
& =\left\vert z\right\vert ^{2}\partial _{\left\vert z\right\vert
^{2}}\log\left\langle z|z\right\rangle _{l}=\left\vert z\right\vert
^{2}h_{l}^{2}\,,
\end{align}%
where the norm can be written as 
\begin{equation}
\left\langle z|z\right\rangle _{l}=\overset{l}{\underset{k=1}{\sum }}%
(f_{l}^{2k-1})^{2}\left\langle z,b_{k}|z,b_{k}\right\rangle +\overset{m}{%
\underset{k=l}{\sum }}(f_{l}^{2k})^{2}\left\langle
z,a_{k}|z,a_{k}\right\rangle \,,
\end{equation}%
So we see that 
\begin{equation}
\left\langle n\right\rangle _{l}=\left\vert z\right\vert ^{2}\partial
_{\left\vert z\right\vert ^{2}}\log\left\langle z|z\right\rangle _{l}\,,
\end{equation}%
which is again a quantity depending on both the position of all the black
rings, and on the ring from which the measurement is being made. The average
occupation number gives the average $U(1)$ $J$ charge of the single trace
part of the operator (\ref{operator_label}), 
\begin{equation}
\left\langle n\right\rangle =\frac{J}{L},\,
\end{equation}%
and also gives the average angular momentum of the string on the $z$-plane.

\subsection{Density Matrix}

\label{denstiy_matrix} 

The states we consider in (\ref{operator_label}) are in general non-BPS
states. They are dual to strings on the bubbling AdS backgrounds described
by systems of concentric rings in the LLM plane. In the situation where we
have only the 1/2-BPS backgrounds, all distinct representations (described
by the Schur polynomials) form an orthonormal basis. Having broken
supersymmetry with the inclusion of the trace operator, such orthonormality
is however no longer guaranteed. Let us consider two very similar background
geometries which have a large number of identical coincident black rings and
a small region where the ring structures differ. The coherent state
wavefunction will be able to distinguish such small differences. The radii,
defining a given ring configuration (\ref{coherent_state_l}), can be viewed
as the continuous parameters in the coherent state string wavefunction, and
we can study the entire class of background geometries from the point of the
view of this wavefunction.

Having obtained the coherent string state on a general background geometry
given by a Young tableau representation $R$, we can write the string state
in this background as 
\begin{equation}
\left\vert z,R\right\rangle \,,
\end{equation}%
where $R$ labels the background geometry, and is seen in the set of
parameters in the coherent state wavefunction (the $R_{i}^{\prime }s$).
There are various constraints we may place on the representation but for now
we study representations with fixed total number of boxes $\Delta $ and $N$
total rows. We would like to define a reduced density matrix by studying the
inner product between different string states in different backgrounds
(representations $R$) and then integrating out the string degrees of
freedom. This can be achieved through the following definition of the
reduced density matrix over coherent states, 
\begin{equation}
\hat{\rho}(R,\hat{R})=\int_{z}\frac{i}{2}dzd\overline{z}\left\langle z,R|z,%
\hat{R}\right\rangle =\int_{z}\left\langle z,R|z,\hat{R}\right\rangle 2\pi
\left\vert z\right\vert d\left\vert z\right\vert \,,
\label{density_matrix_1}
\end{equation}%
where because of the $\theta $ functions in the definition of $\left\langle
z\right\rangle _{l}$ the only contributions to the integral come from the
regions in the LLM plane where the black regions in the representations $R$
and $\hat{R}$ overlap. The wavefunctions are normalized (using the
integration constant $c_{R}$ in (\ref{eq.cl})) such that the reduced density
matrix element of one representation with itself is unity, 
\begin{equation}
\int_{z}\left\langle z,R|z,R\right\rangle 2\pi \left\vert z\right\vert
d\left\vert z\right\vert =1\,.
\end{equation}

Given a particular constraint on the representations (a constraint on the
total number of boxes, rows and columns) we can define an entropy by looking
at all possible inner products between representations within a
superselection sector. The entropy of such a system is given by 
\begin{equation}
S=-Tr(\rho \log \rho )\,,  \label{entropy_information}
\end{equation}%
where $\rho =\frac{1}{D}\hat{\rho},~Tr(\hat{\rho})=D,~Tr(\rho )=1.~$ In
other words we take the trace over the space spanned by the Young tableaux
basis (including the additional Young tableaux plus black inner disk
configurations).

\vspace{1pt}Looking at a fixed total number of boxes and a fixed number of
rows $N$, if we define the number of Young tableaux with no coincident radii
as $D_{1}$, and the total number of Young Young tableaux as $D=D_{1}+D_{2},$
the entropy will be between log$(D_{1})$ and log$(D).$ We can also put a
constraint on the density matrix in terms of the size of the off-diagonal
components. The diagonals of the density matrix are equal, due to the
normalization described above. We define the density matrix as the addition
of the diagonal part and an off-diagonal part $f$,%
\begin{eqnarray}
\rho &=&\frac{1}{D}(I_{D\times D}+f)\,, \\
S &=&\log (D)-\frac{1}{D}Tr((I+f)\log (I+f))\,,  \label{entropy_off_diagonal}
\end{eqnarray}%
where $D$ is the total number of Young tableaux. This density matrix
measures the similarities of different Young tableaux, and due to the
contribution from the off-diagonal components, the entropy gets reduced from
log$(D).$ If the Young tableaux are all identical then the entropy reduces
to zero, and the system becomes a pure state. If on the other hand, the
off-diagonal contributions are small, which is the case for $M\gg N$ (a
sparse distribution of rings with many large gaps), then the entropy is
close to log$(D)$; in that case the second term in (\ref%
{entropy_off_diagonal}) can be approximated as $-\frac{1}{2D}Tr(f^{2})$. We
analyze the dimension $D$ of a constrained set of partitions in Appendix \ref%
{appendix_ dimension_denstiy_matrix}.\vspace{1pt} In general we constrain
the Young tableaux to have fixed $N$ and fixed $M=\frac{\Delta }{2N}$.

\vspace{1pt}In order to calculate the reduced density matrix we must
understand the inner product between two non-equivalent representations.
Consider the two coherent string states in the backgrounds with
representation $R$ and $\hat{R}$ with $m$ and $\hat{m}$ total black rings
respectively. The coherent states are defined as: 
\begin{equation}
\left\vert z,R\right\rangle =\overset{m}{\underset{l=1}{\sum }}\left\{ 
\overset{\infty }{\underset{n=0}{\sum }}\overset{l}{\underset{k=1}{\sum }}%
f_{l}^{2k-1}\left( \frac{{R}_{2k-1}}{z}\right) ^{n}\left\vert
-n,b_{k}\right\rangle +\overset{\infty }{\underset{n=1}{\sum }}\overset{m}{%
\underset{k=l}{\sum }}f_{l}^{2k}\left( \frac{z}{{R}_{2k}}\right)
^{n}\left\vert n,a_{k}\right\rangle \right\} \theta _{{R}_{2l-1},{R}_{2l}}\,
,
\end{equation}

\begin{equation}
\left\vert z,\hat{R}\right\rangle =\overset{{\hat{m}}}{\underset{{\hat{l}}=1}%
{\sum }}\left\{ \overset{\infty }{\underset{n=0}{\sum }}\overset{\hat{l}}{%
\underset{\hat{k}=1}{\sum }}{\hat{f}}_{\hat{l}}^{2\hat{k}-1}\left( \frac{%
\hat{R}_{2\hat{k}-1}}{z}\right) ^{n}\left\vert -n,b_{\hat{k}}\right\rangle +%
\overset{\infty }{\underset{n=1}{\sum }}\overset{m}{\underset{\hat{k}=\hat{l}%
}{\sum }}{\hat{f}}_{\hat{l}}^{2\hat{k}}\left( \frac{z}{{\hat{R}}_{2\hat{k}}}%
\right) ^{{n}}\left\vert n,a_{\hat{k}}\right\rangle \right\} \theta _{\hat{R}%
_{2\hat{l}-1},\hat{R}_{2\hat{l}}}\,.
\end{equation}%
The inner product between the two representations can be written in terms of
a contribution from the inner edges $b$ and outer edges $a$ separately as
there are no cross-terms between $a$'s and $b$'s:%
\begin{equation}
\left\langle z,R|z,\hat{R}\right\rangle =\left\langle z,R|z,\hat{R}%
\right\rangle _{a}+\left\langle z,R|z,\hat{R}\right\rangle _{b}\,.
\end{equation}

Looking at the $a$ type contributions we find the following set of
summations: 
\begin{eqnarray}
\left\langle z,R|z,\hat{R}\right\rangle _{a} &=&\overset{m}{\underset{l=1}{%
\sum }}\overset{{\hat{m}}}{\underset{\hat{l}=1}{\sum }}\overset{m}{\underset{%
k=l}{\sum }}\overset{{\hat{m}}}{\underset{\hat{k}=\hat{l}}{\sum }}\overset{%
\infty }{\underset{n=1}{\sum }}{\hat{f}}_{\hat{l}}^{2\hat{k}}{f}%
_{l}^{2k}\left( \frac{z}{R_{2k}}\right) ^{n}\left( \frac{\overline{z}}{{\hat{%
R}}_{2\hat{k}}}\right) ^{n}\left\langle n,a_{\hat{k}}|n,a_{k}\right\rangle
\theta _{R_{2l-1},R_{2l}}\theta _{\hat{R}_{2\hat{l}-1},\hat{R}_{2\hat{l}}} 
\notag \\
&=&\overset{m}{\underset{l=1}{\sum }}\overset{{\hat{m}}}{\underset{\hat{l}=1}%
{\sum }}\overset{m}{\underset{k=l}{\sum }}\overset{{\hat{m}}}{\underset{\hat{%
k}=\hat{l}}{\sum }}\overset{\infty }{\underset{n=1}{\sum }}{\hat{f}}_{\hat{l}%
}^{2\hat{k}}{f}_{l}^{2k}\left( \frac{\left\vert z\right\vert ^{2}}{%
R_{2k}^{2}-\left\vert z\right\vert ^{2}}\right) \delta _{\hat{R}_{2\hat{k}%
}R_{2k}}\theta _{R_{2l-1},R_{2l}}\theta _{\hat{R}_{2\hat{l}-1},\hat{R}_{2%
\hat{l}}}\,,
\end{eqnarray}%

and similarly for the $b$ type contributions 
\begin{eqnarray}
&&\left\langle z,R|z,\hat{R}\right\rangle _{b}  \notag \\
&=&\overset{m}{\underset{l=1}{\sum }}\overset{{\hat{m}}}{\underset{\hat{l}=1}%
{\sum }}\overset{m}{\underset{k=l}{\sum }}\overset{{\hat{m}}}{\underset{\hat{%
k}=\hat{l}}{\sum }}\overset{\infty }{\underset{n=0}{\sum }}{\hat{f}}_{\hat{l}%
}^{2\hat{k}-1}{f}_{l}^{2k-1}\left( \frac{R_{2k-1}}{z}\right) ^{n}\left( 
\frac{{\hat{R}}_{2\hat{k}-1}}{\overline{z}}\right) ^{n}\left\langle -n,b_{%
\hat{k}}|-n,b_{k}\right\rangle \theta _{R_{2l-1},R_{2l}}\theta _{\hat{R}_{2%
\hat{l}-1},\hat{R}_{2\hat{l}}}  \notag \\
&=&\overset{m}{\underset{l=1}{\sum }}\overset{{\hat{m}}}{\underset{\hat{l}=1}%
{\sum }}\overset{m}{\underset{k=l}{\sum }}\overset{{\hat{m}}}{\underset{\hat{%
k}=\hat{l}}{\sum }}\overset{\infty }{\underset{n=1}{\sum }}{\hat{f}}_{\hat{l}%
}^{2\hat{k}-1}{f}_{l}^{2k-1}\left( -\frac{\left\vert z\right\vert ^{2}}{%
R_{2k-1}^{2}-\left\vert z\right\vert ^{2}}\right) \delta _{\hat{R}_{2\hat{k}%
-1}R_{2k-1}}\theta _{R_{2l-1},R_{2l}}\theta _{\hat{R}_{2\hat{l}-1},\hat{R}_{2%
\hat{l}}}\,,
\end{eqnarray}

The calculation of such an inner product is complicated both by the large
numbers of summations in the above terms, and also by the products implicit
in the $f^{\prime }s$. Having calculated the inner products the integration
must be implemented. The integration itself is not complicated but because
there are very many terms we have performed all such computations using
Mathematica. In order to find the full reduced density matrix it is also
necessary to find all the possible Young tableaux for a given set of
constraints. We also do this using Mathematica. The algorithmic procedure is
outlined in Appendix \ref{appendix_computational_density_matrix} while an
analytical estimate of the dimension of the reduced density matrix is given
in Appendix \ref{appendix_ dimension_denstiy_matrix}. The dimension of the
reduced density matrix for $\Delta =40$, $M=N=9$ is greater than 800
meaning that there are about a million non-equal entries in the density
matrix. Such a computation is possible but, for representations with $M$ and 
$N$ much greater than 10, the combinatorics very quickly become
unmanageable, $D\gtrsim \mathcal{O}(e^{\sqrt{\Delta }})$.

In the following we plot some of the results for various $M$ and $N$, noting
that for $M$ and $N$ of order 10, we can consider that we are approaching
the large $N$ regime.

\begin{figure}[h]
\begin{center}
\includegraphics[width=15cm]{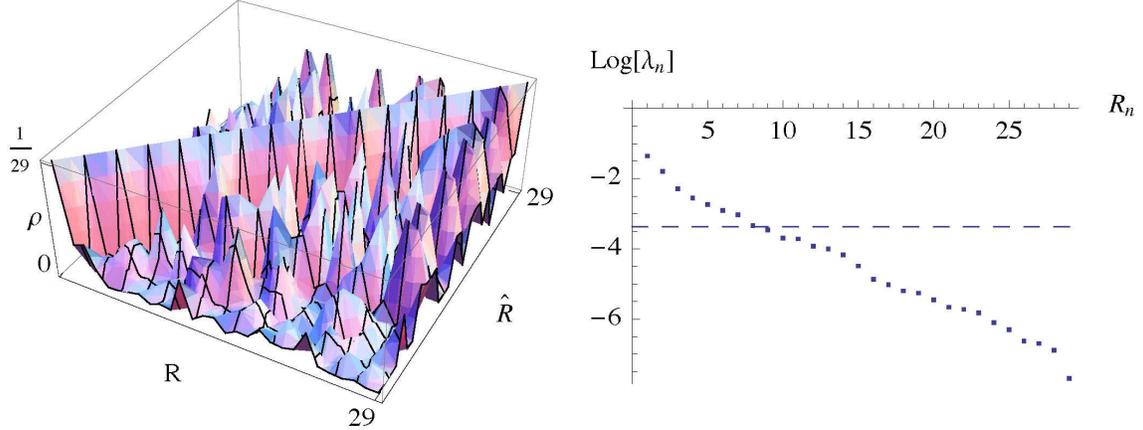}
\end{center}
\caption{{\protect\small The values of the 29 by 29 elements of the reduced
density matrix for the $M=N=6$, $\Delta =18$ ensemble. On the right is the
plot of the logarithm of the 29 eigenvalues of this matrix. The dashed line
is the line that the eigenvalues would fall on if there were no mixing. The
entropy of this particular ensemble is 2.50. The maximum entropy for an
ensemble of 29 elements is $\log 29=3.37$.}}
\label{fig.pl6}
\end{figure}

\begin{figure}[h]
\begin{center}
\includegraphics[width=10cm]{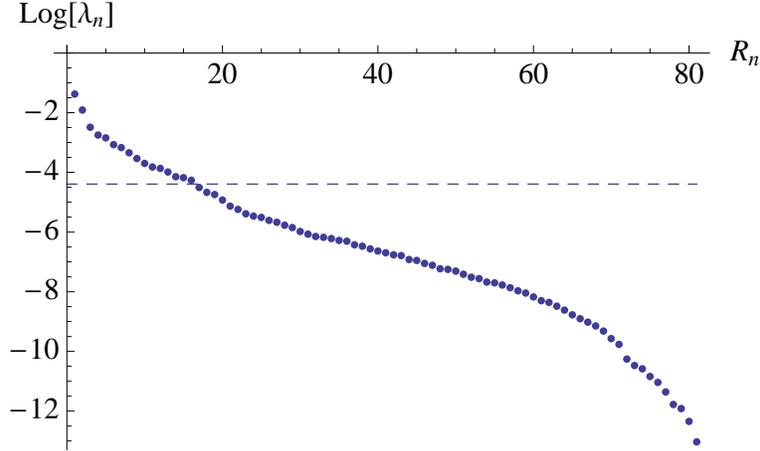}
\end{center}
\caption{{\protect\small The logarithm of the 81 eigenvalues of the reduced
density matrix for the ensemble of representations with $M=N=7$, $\Delta =24$%
. The dashed line corresponds to the positions of the eigenvalues if there
were zero mixing, i.e. $-\log {81}$.}}
\label{fig.p8l}
\end{figure}

\begin{figure}[h]
\begin{center}
\includegraphics[width=12cm]{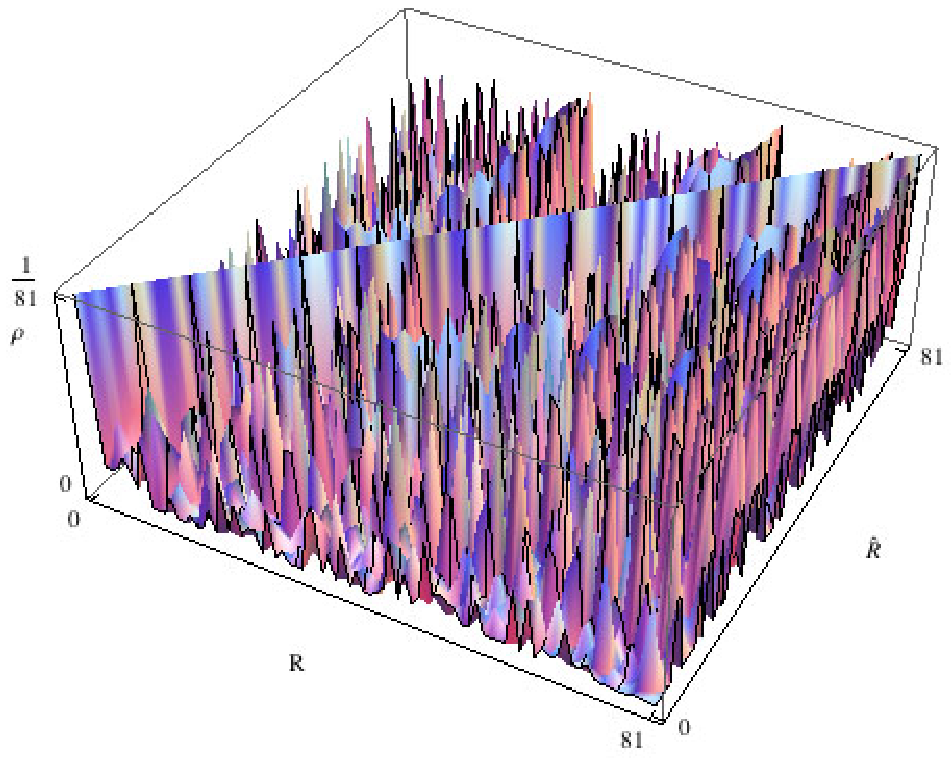}
\end{center}
\caption{{\protect\small The reduced density matrix for $M=N=7$, $\Delta =24$%
. We see the line of $1/81$ on the diagonal and a very large non-trivial
mixing on the off-diagonals.}}
\label{fig.pl8}
\end{figure}

For $M=6,$ $N=6,$ $\Delta =18,$ there are 29 distinct representations (the
representations are given in appendix \ref%
{appendix_computational_density_matrix}). The reduced density matrix can be
diagonalized and the logarithm of the 29 eigenvalues calculated. The
non-diagonalized reduced density matrix along with the logarithm of the
eigenvalues are plotted in figure \ref{fig.pl6}. Note that if there were no
mixing, all eigenvalues would lie on the dashed line. 
Figures \ref{fig.p8l} and \ref{fig.pl8} show the same plots for $M=N=7$, $%
\Delta =24$ for which there are 81 distinct representations. 

\begin{figure}[h]
\begin{center}
\includegraphics[width=10cm]{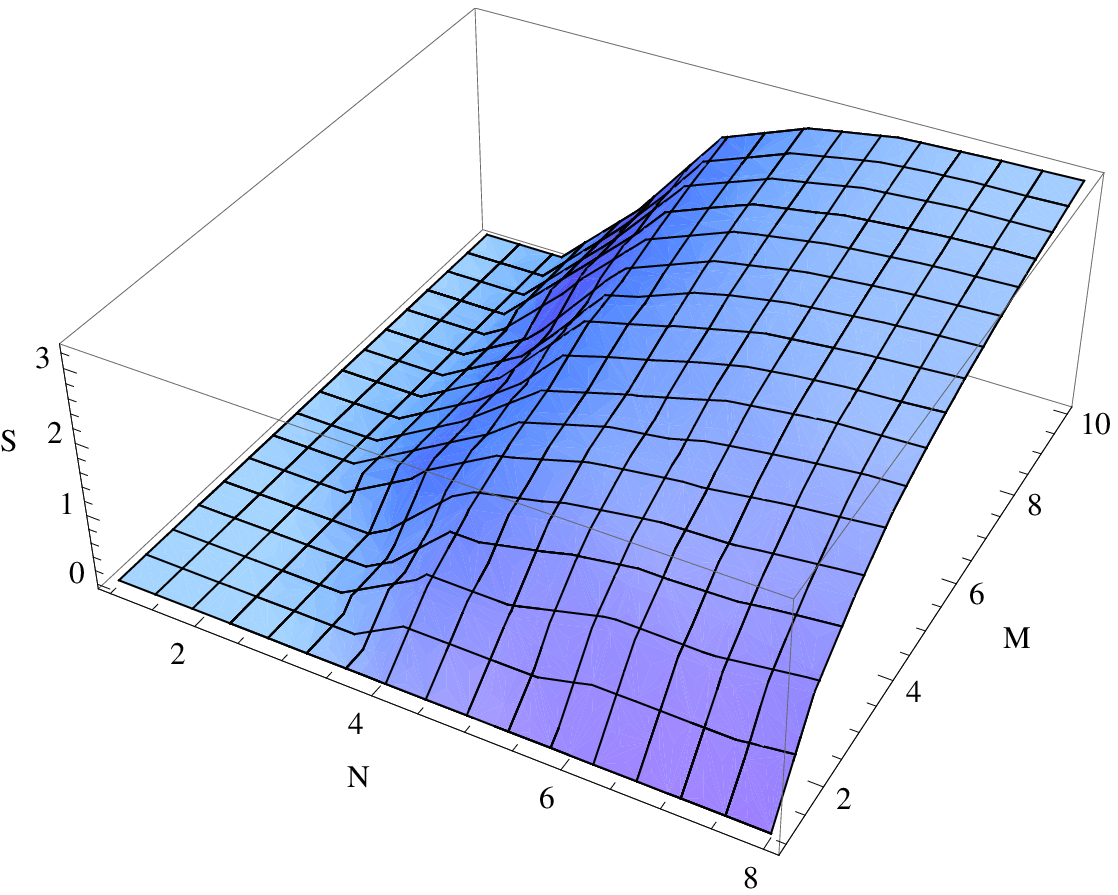}
\end{center}
\caption{{\protect\small Plot of the entropy for the superselection sector $%
\Delta =\frac{MN}{2}$. This plot goes, for large $M$ and $N$ roughly as $%
S\sim\log (M N)\sim \log (\log(D))$. This is a lot slower than the $\log (D) 
$ which would be the answer with minimal mixing between states. The
conclusion therefore is that the non-trivial mixing is extremely important
for $M\sim N$ and $\Delta =\frac{MN}{2}$. Such a conclusion will not be true
for sparse representations.}}
\label{fig.entropyplot}
\end{figure}

In figure \ref{fig.entropyplot} we plot the entropy as a function of $M$ and 
$N$ for $\Delta =\frac{MN}{2}$. 
We see that for large $M$ and $N$ the entropy goes much slower than log$D$
meaning that the off-diagonal elements are extremely important in describing
the purity of the state. Indeed the constraint we are looking at $M=N$ and $%
\Delta =\frac{MN}{2}$ is one in which we would expect a large number of
overlaps in diagrams. This means that, in this case, the number of degrees
of freedom is actually much less than the dimension of the density matrix.

\begin{figure}[h]
\begin{center}
\includegraphics[width=9cm]{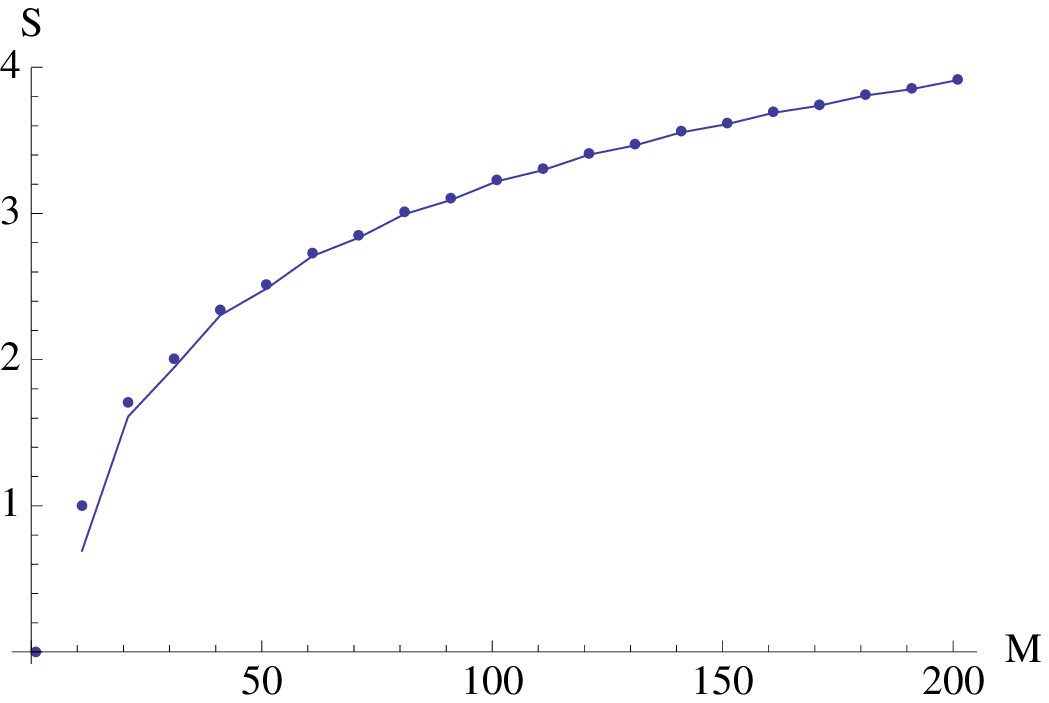}
\end{center}
\caption{{\protect\small Point plot for the entropy as a function of $M$
with $N=3$ and $\Delta =\frac{MN}{2}$ along with the line plot for $\log
(D-1)$ where $D$ is the number of elements in the ensemble for a given $M$,
this value also matches $\log (M/4)$ to extremely high precision. The
conclusion from this is that for $M\gg N$ there is little mixing and the
entropy is maximal.}}
\label{fig.logfit}
\end{figure}

The limit of small $N$, large $M$, however gives a result much closer to $%
\log D$ as can be seen in figure \ref{fig.logfit} where the points are the
values of the entropy calculated for $N=3$ and the line is given by $\log D$%
, showing an almost perfect match in this limit of no overlaps. Note that
there is a subtlety in such a calculation which is dependent on how one
regularizes the divergences in the reduced density matrix. See Appendix \ref%
{appendix_computational_density_matrix} for more details. 

\vspace{1pt}

\vspace{1pt}

\subsection{BMN String States}

\label{BMN_states}

Having studied the continuum limit, we turn to the two-site lattice in the
BMN limit. \vspace{1pt}Such states can be viewed as the two-magnon states of
the string with large $J$ charge. We can project onto states with fixed
angular momentum $J$ by performing the contour integral over the product of
two coherent states as follows: 
\begin{equation}
\left\vert \phi ^{J}\right\rangle =\oint_{C}dzz^{-(J+1)}\left\vert
z\right\rangle \left\vert z\right\rangle \,,
\end{equation}%
however unlike in the one black-ring case \cite{Chen:2007gh}, $\left\vert
z\right\rangle $ is a sum of contributions from each ring, so we have to sum
over rings: 
\begin{equation}
\left\vert z\right\rangle \left\vert z\right\rangle
=\sum_{l=1}^{m}\sum_{l^{\prime }=1}^{m}\left\vert z\right\rangle
_{l}\left\vert z\right\rangle _{l^{\prime }}\theta (l)\theta (l^{\prime })\,,
\end{equation}%
where $\theta (l)$ is the theta function with unit support only on the $%
l^{th}$ ring (this is simply a rewriting of (\ref{eq.z}) without referring
to the ring radii).

Since we are only looking at a single representation, $R$, (both $\left\vert
z\right\rangle ^{\prime }$s are the same), the product of $\theta ^{\prime
}s $ means that the above simplifies to: 
\begin{equation}
\left\vert z\right\rangle \left\vert z\right\rangle
=\sum_{l=1}^{m}\left\vert z\right\rangle _{l}\left\vert z\right\rangle
_{l}\theta (l)\,.
\end{equation}%
We therefore start by looking at $\left\vert z\right\rangle _{l}\left\vert
z\right\rangle _{l}$ 
\begin{eqnarray}
\left\vert \phi ^{J}\right\rangle _{l} &=&\oint_{C}dzz^{-(J+1)}\left\vert
z\right\rangle _{l}\left\vert z\right\rangle _{l}  \notag \\
&=&\oint_{C}dzz^{-(J+1)}\left( \sum_{k=1}^{l}\sum_{n=0}^{\infty
}f_{l}^{2k-1}\left( \frac{R_{2k-1}}{z}\right) ^{n}\left\vert
-n,b_{k}\right\rangle +\sum_{k=l}^{m}\sum_{n=1}^{\infty }f_{l}^{2k}\left( 
\frac{z}{R_{2k}}\right) ^{n}\left\vert n,a_{k}\right\rangle \right)  \notag
\\
&&\cdot \left( \sum_{p=1}^{l}\sum_{s=0}^{\infty }f_{l}^{2p-1}\left( \frac{%
R_{2p-1}}{z}\right) ^{s}\left\vert -s,b_{p}\right\rangle
+\sum_{p=l}^{m}\sum_{s=1}^{\infty }f_{l}^{2p}\left( \frac{z}{R_{2p}}\right)
^{s}\left\vert s,a_{p}\right\rangle \right) \,.
\end{eqnarray}%
Only terms which have a single negative power of $z$ will contribute to the
integrand. Therefore, for positive $J$ this is only possible for terms of
the form $\left\vert -n\right\rangle \left\vert n+J\right\rangle $ , $%
\left\vert n+J\right\rangle \left\vert -n\right\rangle $, $\left\vert
n\right\rangle \left\vert J-n\right\rangle $ and $\left\vert
J-n\right\rangle \left\vert n\right\rangle $. There will be no contributions
from a product of two negative movers (two $b$ types) only.

Calculating the contribution from each $l$ and summing we arrive at the
final expression for the angular momentum eigenstate as: 
\begin{eqnarray}
\left\vert \phi ^{J}\right\rangle &=&\sum_{l=1}^{m}\left\{ \left(
\sum_{k=1}^{l}\sum_{n=0}^{\infty
}\sum_{p=l}^{m}f_{l}^{2k-1}f_{l}^{2p}R_{2k-1}^{n}R_{2p}^{-(n+J)}\left\vert
-n,b_{k}\right\rangle \left\vert n+J,a_{p}\right\rangle \right) \right. 
\notag \\
&+&\left( \sum_{k=l}^{m}\sum_{n=0}^{\infty
}\sum_{p=1}^{l}f_{l}^{2k}f_{l}^{2p-1}R_{2k}^{-(n+J)}R_{2p-1}^{n}\left\vert
n+J,a_{k}\right\rangle \left\vert -n,b_{p}\right\rangle \right)  \notag \\
&+&\left. \left(
\sum_{k=l}^{m}\sum_{n=1}^{J-1}%
\sum_{p=l}^{m}f_{l}^{2k}f_{l}^{2p}R_{2k}^{-n}R_{2p}^{n-J}\left\vert
n,a_{k}\right\rangle \left\vert J-n,a_{p}\right\rangle \right) \right\} \,.
\label{state_angular_momemtum}
\end{eqnarray}%
\vspace{1pt}

The analysis for the total negative $J$ states is similar, except that we
now pick up products of two $b$ type states, and do not pick up the products
of two $a$ type states.

We can now use (\ref{state_angular_momemtum}) as a general ansatz to write
down an energy eigenstate. Simplifying this expression we arrive at, 
\begin{eqnarray}
\left\vert \phi ^{J}\right\rangle &=&\sum_{k=1}^{m}\sum_{p=1}^{m}\left\{
\left( \sum_{n=0}^{\infty
}F_{-n}^{2k-1,2p}R_{2k-1}^{n}R_{2p}^{-(n+J)}\left\vert -n,n+J\right\rangle
_{k,p}\right. \right.  \notag  \label{state_angular_momentum_2} \\
&+&\left. F_{n+J}^{2k,2p-1}R_{2k}^{-(n+J)}R_{2p-1}^{n}\left\vert
n+J,-n\right\rangle _{k,p}\right) +\left. \left(
\sum_{n=1}^{J-1}F_{n}^{2k,2p}R_{2k}^{-n}R_{2p}^{n-J}\left\vert
n,J-n\right\rangle _{k,p}\right) \right\} \,,  \notag \\
&&
\end{eqnarray}%

where we write the two occupation numbers together into one bracket and
leave the information about which ring is being excited as a subsript after
the ket. The $F^{\prime }s$ are calculated after applying an energy
eigenstate constraint on the kets. A full explanation of this procedure can
be found in Appendix \ref{appendix_BMN_states}. In order to find the correct
constraints on the $F^{\prime }s$ we have to simplify the problem by looking
at a decoupling limit which reduces us to the two site case already
discussed extensively in \cite{Chen:2007gh}. The full problem of finding the 
$F^{\prime }s$ in complete generality is outside the scope of the current
paper.

\vspace{1pt}The decoupling limit which we choose is the one in which a
single black ring, the $l^{th}$ is decoupled from the rest. In (\ref%
{state_angular_momentum_2}), we pick $k=p=l,$ and since $R_{2l-1}= \sqrt{%
C_{b_{l}}/N}$ and $R_{2l}=\sqrt{C_{a_{l}}/N}$ we can define $\gamma
=C_{b_{l}}/{N_{l}},$ and the state and algebra are very similar to the one
black ring case \cite{Chen:2007gh}. From the general expression (\ref%
{state_angular_momentum_2}), we get in the decoupling limit that 
\begin{equation}
\left\vert \phi ^{J}\right\rangle =\sum_{n=-\infty }^{0}\left( \frac{\gamma 
}{1+\gamma }\right) ^{-\frac{n}{2}}f_{n}v_{n}+\sum_{n=1}^{J-1}f_{n}v_{n}+%
\sum_{n=J}^{\infty }\left( \frac{\gamma }{1+\gamma }\right) ^{\frac{n-J}{2}%
}f_{n}v_{n}\, ,
\end{equation}%
where $v_{n}=\left\vert J-n,n\right\rangle $.

The energy eigenstate equation is 
\begin{equation}
H\left\vert \phi \right\rangle =2\lambda (a_{1}^{\dagger
}a_{1}+a_{2}^{\dagger }a_{2}-a_{2}^{\dagger }a_{1}-a_{1}^{\dagger
}a_{2})\left\vert \phi \right\rangle =E\left\vert \phi \right\rangle 
\end{equation}%
i.e. 
\begin{equation}
\left( 2\lambda (a_{1}^{\dagger }a_{1}+a_{2}^{\dagger }a_{2}-a_{2}^{\dagger
}a_{1}-a_{1}^{\dagger }a_{2})-E\right) \left\vert \phi \right\rangle =0\,.  \label{eigenstate_equation}
\end{equation}

Applying this operation to our state we get: 
\begin{eqnarray}
&&2\lambda \left\{ \sum_{n=-\infty }^{0}\left( \frac{\gamma }{1+\gamma }%
\right) ^{-\frac{n}{2}}f_{n}\left( (1+2\gamma )v_{n}-\sqrt{\gamma (1+\gamma )%
}(v_{n-1}+v_{n+1})\right) \right.  \notag \\
&+&\sum_{n=1}^{J-1}f_{n}\left( (2+2\gamma v_{n}-(1+\gamma
)(v_{n-1}+v_{n+1})\right)  \notag \\
&+&\left. \sum_{n=J}^{\infty }\left( \frac{\gamma }{1+\gamma }\right) ^{%
\frac{n-J}{2}}f_{n}\left( (1+2\gamma )v_{n}-\sqrt{\gamma (1+\gamma )}%
(v_{n-1}+v_{n+1})\right) \right\}  \notag \\
&-&\frac{N}{N_{l}}E\left\{ \sum_{n=-\infty }^{0}\left( \frac{\gamma }{%
1+\gamma }\right) ^{-\frac{n}{2}}f_{n}v_{n}+\sum_{n=1}^{J-1}f_{n}v_{n}+%
\sum_{n=J}^{\infty }\left( \frac{\gamma }{1+\gamma }\right) ^{\frac{n-J}{2}%
}f_{n}v_{n}\right\} =0\,.  \notag \\
&&  \label{generating_recursive_equations}
\end{eqnarray}

This equation generates recursion relations between triples of $%
f_{n}{}^{\prime }s$ and we can solve the whole system iteratively starting
from just two $f_{n}{}^{\prime }s,$ e.g. $f_{0},f_{1}$. These recursive
equations will be almost the same as \cite{Chen:2007gh}, except that in (\ref%
{generating_recursive_equations}) we have ${\hat{E}}=\frac{N}{N_{l}}E$
instead of $E$. \ We thus have three unknowns, $f_{0},f_{1},{\hat{E}.~}$
$f_{0}$ is set to 1 via an overall normalization freedom while $f_{1}$ and $%
\hat{E}$ are fixed by demanding convergence of the series in the two
asymptotic regimes $n\rightarrow \pm \infty $. The calculation is discussed
in detail in the appendix but the outline is the following: Three generating
functions are derived which describe in closed form the recursion relations
between $f_{0}$ and $f_{J-1}$, $f_{J+1}$ and $f_{\infty }$ and, $f_{-1}$ and 
$f_{-\infty }$. Using these generating functionals we are able to write $%
f_{-\infty }$ and $f_{\infty }$ in as functions of $\hat{E}$ and $f_{1}$.
Convergence of the sums puts a constraint first on $f_{1}$ from the $%
n\rightarrow \infty $ regime and finally a constraint equation is found for $%
\hat{E}$ from the $n\rightarrow -\infty $ regime. More detailed discussions
are in Appendix \ref{appendix_BMN_states}. One finally gets that for large $%
J $ states, the eigenvalues are \footnote{
The state $\left\vert n_{1},n_{2},...,n_{L}\right\rangle$ is equivalent to the state under a cyclic permutation of the integers, e.g. to $\left\vert n_{L},n_{	1},...,n_{L-1} \right\rangle$ \cite{Chen:2007gh}, due to the cyclicity of the trace part of the operator (\ref{operator_label}). For a two-site system, one can write a general state as the sum of a symmetric state, i.e. one with the property $\left\vert n_{1},n_{2} \right\rangle=\left\vert n_{2},n_{1} \right\rangle$, and an anti-symmetric state, i.e. one with the property $\left\vert n_{1},n_{2} \right\rangle=-\left\vert n_{2},n_{1} \right\rangle$. The energy eigenstate equation (\ref{eigenstate_equation}) is to be supplemented by the projection onto symmetric eigenstates. Thereby one should project out half of the eigenfunctions of the equation, associated with ${\hat n}=2n+1$, and only keep the other half with ${\hat n}=2n$ \cite{Chen:2007gh}, and interpret the $n$ as the mode number of the BMN state. Therefore finally (\ref{spectrum_positive_J}), (\ref{spectrum_negative_J}) are in agreement with \cite{Chen:2007gh}.
} 
\begin{equation}
E_{ { \hat n}}=\frac{C_{a_{l}}}{N}\frac{2\pi ^{2} {\hat n}^{2}\lambda }{J^{2}}\left( 1-\frac{%
2+4\gamma }{J}\right) \,   =\frac{C_{a_{l}}}{N}\frac{8\pi ^{2} n^{2}\lambda }{J^{2}}\left( 1-\frac{%
2+4\gamma }{J}\right) \,  ,    \label{spectrum_positive_J}
\end{equation}%
where here ${ \hat n}=2n$, and $n$ is the oscillator mode number in the BMN state \cite%
{Berenstein:2002jq}. A similar procedure for large $-J$ states produces
similar recursive relations and yields

\begin{equation}
E_{{ n}}=\frac{C_{b_{l}}}{N}\frac{8\pi ^{2}{n}^{2}\lambda }{J^{2}}\left( 1-\frac{%
2+4\gamma }{J}\right)\, .   \label{spectrum_negative_J}
\end{equation}%
These are strings excited along the exterior and interior edges of the black ring. The different coefficients in the energy spectrum reflect the different
effective radii of curvature when we take the plane-wave limit along the
geodesics near the different circles \cite{Lin:2004nb}. Thus, by measuring
and comparing the effective units of energies of the states, one can know
the exact location of these (short) string excitations in the bulk spacetime.

\section{Discussion}

\label{discussion}

In this paper we have studied the quantum string excitations on
top of a class of axially symmetric bubbling geometries dual to the 1/2 BPS
chiral primary operators. This work is a continuation of the research
undertaken in \cite{Vazquez:2006id},\cite{Chen:2007gh},\cite{Koch:2008ah},%
\cite{Koch:2008cm}. On the gauge theory side, we have studied operators of
the form of a product of a Schur polynomial $\chi _{R}(\mathcal{Z})$ and a
single trace of $\mathcal{Z}^{\prime }s$ and impurities $Y^{\prime }s$. The
Schur polynomial part corresponds to a general Young tableau which specifies
a background geometry. The single trace part can be viewed as a bosonic
lattice of $Y^{\prime }s,$ and for each site there are, in general, $2m$
type of bosonic excitations, $m$ of which have positive occupation numbers $%
n $, corresponding to the $m$ inward corners of the Young tableau, at which
point one can add $n$ extra boxes there. The other $m$, with negative
occupation numbers, $-n$, correspond to the $m$ outward corners on the Young
tableau, where one can subtract $n$ boxes.

Developing the work of \cite{Vazquez:2006id},\cite{Chen:2007gh},\cite%
{Koch:2008ah} we have solved the problem of the coherent state for a general
geometry with arbitrary concentric ring distributions (i.e. a general set of
radii), and equivalently for a general Young tableau. \vspace{1pt}We have
also studied the algebra of the shift operators (or ladder operators) $%
a,a^{\dagger }$, acting on the different types of bosonic excitations on the
lattice, and extended the algebra of \cite{Koch:2008ah} by including the
mixing of the $\left\vert 0\right\rangle $ states of different black rings.
These different types of excitation correspond to the different possible
circles in the geometry. The coherent state wavefunction has different
expressions on the $m$ different and disconnected black rings. On each black
ring, the contribution to the Fock state wavefunction, normalizable on that
ring, comes from all the positive mover states along the circles outside
that ring, and all the negative mover states along the circles inside that
ring.

The norm of the coherent state wavefunction is related to the metric
function $V$ that appears in the dual string sigma model \cite%
{Vazquez:2006id},\cite{Chen:2007gh},\cite{Koch:2008ah}. The radii that
appear in the denominator of the norm are those radii for which the angular
momentum excitations of the string are included in the coherent state
expansion. The radii that appear in the numerator of the norm are all the
rest of the radii in the geometry, and emerge due to the factorization of
the numerator after summation. We have also calculated the average
occupation number on each black ring, which are related to the metric
function $V$. We have checked that the coherent state wavefunction behaves
nicely under the merging of two nearby rings.

Having obtained the coherent state wavefunction for a general concentric
ring geometry, or equivalently, for a general Young tableau, we used the
wavefunction as a means to study the different background geometries since
the continuous parameters of these geometries are encoded in the coherent
state wavefunction itself. We have then gone on to define a reduced density
matrix element between any pair of geometries or Young tableaux. This has
been implemented by integrating out the string degrees of freedom from the
cross product between the coherent state wavefunctions living on these
geometries. The reduced density matrix measures the degree of similarity
between any pair of geometries. In the most extreme case, if the two
geometries are almost identical except for some very small inequivalent
region, the matrix element between them is close to $1/D$, where $D$ is the
total number of Young tableaux in a given constrained ensemble. We use
computational methods to obtain both the representations in the ensemble and
furthermore the reduced density matrix itself. This calculation is a novel
development, which allows us to study the mixing of non-supersymmetric
operators. The mixing of the tableaux is due to the non-BPS property of the
entire operator (Schur polynomial times single trace).

We have also studied an ansatz similar to the coherent states that describe
the BMN states with large $J$ charge. The coherent state has no fixed total $%
J$, but includes a superposition of all states with different $J^{\prime }s$%
. By picking the fixed $J$ solutions from within the coherent state
expansion, and using these states as the basis for the superposition to form
a BMN eigenstate of the dilatation generator, we are able to solve the
eigenvalue problem and find the eigenvalues. The spectrums encode the
different radii of the rings within the geometry and therefore the locations
of short string excitations in the bulk gravity dual. It therefore tells us
some locality information of the bulk quantum gravity. We also show that one can
obtain the BMN states with large negative $J$ charge, which correspond to
the negative movers along the inner circles of the different rings. For
fixed large $|J|$, the number of sites of the lattice $L$ corresponds to the
number of magnons on the string.

In finding the string sigma model in these geometric backgrounds we obtain a
metric function, $V$ , which describes the entire class of concentric ring
geometries. The metric function $V$ is a superposition of many terms each of
which looks like a metric function for an $AdS_{5}$ space with a different
radius. A future goal would be to consider the solutions, and S-matrices of
these sigma models in more detail. The straight string solutions are of
particular interest, and may be related to the work of \cite{Hofman:2006xt}, 
\cite{Beisert:2006qh}. Semi-classical long strings on these backgrounds were
studied previously in \cite{Filev:2004yv}. It would seem natural to consider
these as operators of the form of the product of the Schur polynomial and an
additional operator.

A further direction of research would be to consider the higher loop dilatation
generators acting on the lattice, having here extensively studied the
application of the one loop dilatation generator. It would also be
interesting to generalize the trace operator under consideration from one
which contains only $\mathcal{Z}$ and $Y$ impurities to one which contains
fields from a larger sector, for instance the derivative operator. There is
still much to be done to fully understand many aspects of these types of
operators.

\vspace{1pt}

\section*{Acknowledgments}

This work is supported in part by the ME and Feder (grant FPA2008- 01838),
by the Spanish Consolider-Ingenio 2010 Programme CPAN (CSD2007-00042), by
the Juan de la Cierva program of MCyI of Spain, and by Xunta de Galicia
(Conselleria de Educacion and grants PGIDIT06 PXIB206185Pz and INCITE09 206
121 PR). H.L. also would like to thank Bin Chen, Bo Feng, Oleg Lunin, Juan
Maldacena for some correspondences, and University of Chicago, McGill
University and CERN for hospitalities. AM is supported by FCT grant SFRH/BPD/65058/2009. A.M. also would like to thank the University of Santiago de Compostela for their hospitality.

\appendix

\renewcommand{\theequation}{A.\arabic{equation}} \setcounter{equation}{0}

\section{Derivation of the coherent states}

\label{appendix_coherent_state_derivation}

Here we give a more detailed explanation of the calculation of
the coherent state, used extensively throughout this work.

We assume that the coherent state on the $l^{th}$ ring takes the form

\begin{equation}
\left\vert z\right\rangle _{l}=\overset{m}{\underset{k=1}{\sum }}%
g_{l}^{k}\left\vert 0\right\rangle _{k}+\overset{l}{\underset{k=1}{\sum }}%
f_{l}^{2k-1}\overset{\infty }{\underset{n=1}{\sum }}\left( \frac{R_{2k-1}}{z}%
\right) ^{n}\left\vert -n,b_{k}\right\rangle +\overset{m}{\underset{k=l}{%
\sum }}f_{l}^{2k}\overset{\infty }{\underset{n=1}{\sum }}\left( \frac{z}{%
R_{2k}}\right) ^{n}\left\vert n,a_{k}\right\rangle \,,
\end{equation}%
where the excitations on the inner and outer rings are labelled by $b_{k}$
and $a_{k}$ and the ground state of the $k^{th}$ ring is labelled by $%
\left\vert 0\right\rangle _{k}$. The ranges of the sum over $k$ is due to
the normalizability on $R_{2l-1}<|z|<R_{2l}$.

The algebras are written in section \ref{operators}, and in particular, we
assume that the action of the annihilation operators on a single positive
mover gives a sum over zero modes: 
\begin{equation}
a\left\vert 1,a_{l}\right\rangle =\overset{m}{\underset{k=1}{\sum }}v_{l}^{k}%
\frac{\sqrt{C_{a_{k}}}}{\sqrt{N}}\left\vert 0\right\rangle _{k},~\ \ \ \
~a\left\vert 0\right\rangle _{k}=\frac{\sqrt{C_{b_{k}}}}{\sqrt{N}}\left\vert
-1,b_{k}\right\rangle \,.\qquad
\end{equation}%
The mixing of the $\left\vert 0\right\rangle _{k}$ states is natural, since
when two nearby black rings merge, or equivalently, two nearby vertical
edges of the Young tableau merge together, the $\left\vert 0\right\rangle $
states on these two black rings, under this limit, have to be identified as
the same state. The algebra without such mixing would reduce to $m$
decoupled algebras, each of which has a similar form but with different
coefficients.

We then have that the action of the annihilation operator on the negative
and positive movers respectively gives%
\begin{align}
a\overset{\infty }{\underset{n=1}{\sum }}\left( \frac{R_{2k-1}}{z}\right)
^{n}\left\vert -n,b_{k}\right\rangle & =\overset{\infty }{\underset{n=1}{%
\sum }}\left( \frac{R_{2k-1}}{z}\right) ^{n}\frac{\sqrt{C_{b_{k}}}}{\sqrt{N}}%
\left\vert -n-1,b_{k}\right\rangle  \notag \\
& =z\overset{\infty }{\underset{n=1}{\sum }}\left( \frac{R_{2k-1}}{z}\right)
^{n}\left\vert -n,b_{k}\right\rangle -\frac{\sqrt{C_{b_{k}}}}{\sqrt{N}}%
\left\vert -1,b_{k}\right\rangle\, ,
\end{align}

\begin{align}
a\overset{\infty }{\underset{n=1}{\sum }}\left( \frac{z}{R_{2k}}\right)
^{n}\left\vert n,a_{k}\right\rangle & =\overset{\infty }{\underset{n=2}{\sum 
}}\left( \frac{z}{R_{2k}}\right) ^{n}\frac{\sqrt{C_{a_{k}}}}{\sqrt{N}}%
\left\vert n-1,a_{k}\right\rangle +\frac{z}{R_{2k}}\overset{m}{\underset{j=1}%
{\sum }}v_{k}^{j}\frac{\sqrt{C_{a_{k}}}}{\sqrt{N}}\left\vert 0\right\rangle
_{j}  \notag \\
& =z\overset{\infty }{\underset{n=1}{\sum }}\left( \frac{z}{R_{2k}}\right)
^{n}\left\vert n,a_{k}\right\rangle +z\overset{m}{\underset{j=1}{\sum }}%
v_{k}^{j}\left\vert 0\right\rangle _{j}\,.
\end{align}%
The coherent state condition then needs to be satisfied which will provide a
series of recursion relations for the $f^{\prime }s$. The condition 
\begin{equation}
a\left\vert z\right\rangle _{l}-z\left\vert z\right\rangle _{l}=0\,
\end{equation}%
yields the constraint%
\begin{equation}
\overset{m}{\underset{k=1}{\sum }}g_{l}^{k}\frac{\sqrt{C_{b_{k}}}}{\sqrt{N}}%
\left\vert -1,b_{k}\right\rangle -\overset{l}{\underset{k=1}{\sum }}%
f_{l}^{2k-1}\frac{\sqrt{C_{b_{k}}}}{\sqrt{N}}\left\vert
-1,b_{k}\right\rangle +\overset{m}{\underset{k=l}{\sum }}f_{l}^{2k}(\overset{%
m}{\underset{j=1}{\sum }}v_{k}^{j}z\left\vert 0\right\rangle _{j})-\overset{m%
}{\underset{k=1}{\sum }}g_{l}^{k}z\left\vert 0\right\rangle _{k}=0\,.
\end{equation}%
which gives a relationship between the $f^{\prime }s$ and $g^{\prime }s$ 
\begin{equation}
g_{l}^{k}=f_{l}^{2k-1},\quad k\leqslant l;\qquad g_{l}^{k}=0,\quad k>l\,.
\end{equation}%
We also have the following constraint coming from the sums over zero modes 
\begin{equation}
\overset{m}{\underset{k=l}{\sum }}f_{l}^{2k}(\overset{m}{\underset{j=1}{\sum 
}}v_{k}^{j}\left\vert 0\right\rangle _{j})=\overset{l}{\underset{j=1}{\sum }}%
f_{l}^{2j-1}\left\vert 0\right\rangle _{j}\,,
\end{equation}%
which can be rewritten as 
\begin{equation}
\overset{m}{\underset{k=l}{\sum }}f_{l}^{2k}v_{k}^{j}=f_{l}^{2j-1},~~j%
\leqslant l;~~~~~\overset{m}{\underset{k=l}{\sum }}%
f_{l}^{2k}v_{k}^{j}=0,~~j>l\,,
\end{equation}%
and thus%
\begin{equation}
v_{l}^{j}=\frac{\bar{v}_{l}^{j}}{f_{l}^{2l}},~~~\bar{v}_{l}^{j}=f_{l}^{2j-1}-%
\overset{m}{\underset{k=l+1}{\sum }}\frac{f_{l}^{2k}}{f_{k}^{2k}}\bar{v}%
_{k}^{j},~~~~\bar{v}_{m}^{j}=f_{m}^{2j-1},~~\ \bar{v}%
_{m-1}^{j}=f_{m-1}^{2j-1}-\frac{f_{m-1}^{2m}}{f_{m}^{2m}}f_{m}^{2j-1}\,.
\end{equation}

Taking the above constraints we are left with a simplified expression for
the coherent state, given by:%
\begin{equation}
\left\vert z\right\rangle _{l}=\overset{l}{\underset{k=1}{\sum }}%
f_{l,m}^{2k-1}\overset{\infty }{\underset{n=0}{\sum }}\left( \frac{R_{2k-1}}{%
z}\right) ^{n}\left\vert -n,b_{k}\right\rangle +\overset{m}{\underset{k=l}{%
\sum }}f_{l,m}^{2k}\overset{\infty }{\underset{n=1}{\sum }}\left( \frac{z}{%
R_{2k}}\right) ^{n}\left\vert n,a_{k}\right\rangle \,,
\end{equation}%
where we have made a choice to absorb the zero modes in the $b_{k}$ states.
From the coherent state expansion we get the norm of the state on the $%
l^{th} $ ring as%
\begin{eqnarray}
\left\langle z|z\right\rangle _{l} &=&\overset{l}{\underset{k=1}{\sum }}%
\frac{c_{l,m}^{2k-1}R_{2k-1}^{2}}{R_{2k-1}^{2}-\left\vert z\right\vert ^{2}}+%
\overset{m}{\underset{k=l}{\sum }}\frac{c_{l,m}^{2k}R_{2k}^{2}}{%
R_{2k}^{2}-\left\vert z\right\vert ^{2}}, \\
~c_{l,m}^{2k-1} &=&-(f_{l,m}^{2k-1})^{2},~\
~~~~c_{l,m}^{2k}=(f_{l,m}^{2k})^{2}\,,
\end{eqnarray}%
which can be re-summed and factored in terms of a quotient of products as 
\begin{equation}
\left\langle z|z\right\rangle _{l}=c_{l}\frac{\overset{l-1}{\underset{k=1}{%
\prod }}\left( R_{2k}^{2}-\left\vert z\right\vert ^{2}\right) \overset{m}{%
\underset{k=l+1}{\prod }}\left( R_{2k-1}^{2}-\left\vert z\right\vert
^{2}\right) (-\left\vert z\right\vert ^{2})}{\overset{l}{\underset{k=1}{%
\prod }}\left( R_{2k-1}^{2}-\left\vert z\right\vert ^{2}\right) \overset{m}{%
\underset{k=l}{\prod }}\left( R_{2k}^{2}-\left\vert z\right\vert ^{2}\right) 
}\,,
\end{equation}%
where $c_{l}$ is an overall normalization constant.

Since when there is no black region in the center, the norm is zero when $z=0$:
\begin{equation}
\overset{l}{\underset{k=1}{\sum }}c_{l,m}^{2k-1}+\overset{m}{\underset{k=l}{%
\sum }}c_{l,m}^{2k}=0\,,
\end{equation}%
which has solution 
\begin{eqnarray}
(f_{l,m}^{2k-1})^{2} &=&\frac{-c_{l}%
\prod_{p=l+1}^{m}(R_{2k-1}^{2}-R_{2p-1}^{2})}{\prod_{p=1\neq
k}^{l}(R_{2k-1}^{2}-R_{2p-1}^{2})}\frac{%
\prod_{p=1}^{l-1}(R_{2k-1}^{2}-R_{2p}^{2})}{%
\prod_{p=l}^{m}(R_{2k-1}^{2}-R_{2p}^{2})} \\
&=&-c_{l}\prod_{p=1\neq k}^{m}\left( R_{2k-1}^{2}-R_{2p-1}^{2}\right) ^{%
\text{sign}(p-l)}\prod_{p=1}^{m}\left( R_{2k-1}^{2}-R_{2p}^{2}\right) ^{%
\text{sign}(l-p-1)}\,
\end{eqnarray}%
\begin{eqnarray}
(f_{l,m}^{2k})^{2} &=&\frac{c_{l}\prod_{p=l+1}^{m}(R_{2k}^{2}-R_{2p-1}^{2})}{%
\prod_{p=1}^{l}(R_{2k}^{2}-R_{2p-1}^{2})}\frac{\prod_{p=1\neq
k}^{l-1}(R_{2k}^{2}-R_{2p}^{2})}{\prod_{p=l\neq k}^{m}(R_{2k}^{2}-R_{2p}^{2})%
} \\
&=&c_{l}\prod_{p=1}^{m}\left( R_{2k}^{2}-R_{2p-1}^{2})\right) ^{\text{sign}%
(p-l)}\prod_{p=1\neq k}^{m}\left( R_{2k}^{2}-R_{2p}^{2}\right) ^{\text{sign}%
(l-p-1)}\,,
\end{eqnarray}%
where sign$(p-l)=1,$ for $p\geqslant l$; and sign$(p-l)=-1,$ for $p<l$.

It can be seen that 
\begin{eqnarray}
\text{sign}(c_{l,m}^{2k-1}) &=&(-1)^{2(m-k)+1}=(-1)^{\#R>R_{2k-1}}=-1\,, \\
\text{sign}(c_{l,m}^{2k}) &=&(-1)^{2(m-k)}=(-1)^{\#R>R_{2k}}=1\,.
\end{eqnarray}%
The $f_{l}^{2k-1}$ and $f_{l}^{2k}$ coefficients give the relative strengths
of the response of the $l^{th}$ ring to the excitations on $R_{2k-1}$,$%
R_{2k} $ circles in the $m$-black-ring geometry.

It can be checked that they correctly produce all the non-zero coefficients
of the 2-circle case with 
\begin{equation}
c_{1,1}=1=R_{2}^{2}-R_{1}^{2}\, ,
\end{equation}%
since $R_{2}^{2}-R_{1}^{2}$ is already proportional to $N$ and we use the
coordinate $z$ in the units that $R_{2}^{2}-R_{1}^{2}$ is 1.

It can also be checked that they correctly produce all the non-zero
coefficients of the 3-circle case with 
\begin{equation}
c_{1,2}=\frac{R_{4}^{2}R_{2}^{2}}{R_{3}^{2}},\qquad c_{2,2}=\frac{%
R_{4}^{2}R_{3}^{2}}{R_{2}^{2}}\,.
\end{equation}%
These match with the 3-circle result in \cite{Chen:2007gh}\footnote{%
There is a slight typo in equation (58) of \cite{Chen:2007gh}, where the
prefactors of the last two terms are switched.}. Note our 4,3,2 is their
3,2,1, we have taken the $R_{1}^{2}=0$ limit, and these normalizations make
the norm of coherent state when $R_{1}^{2}=0,$ to be 1 (in the unit of
$c_R$) at the origin, $z=0$. In this procedure, we first take $R_{1}^{2}=0,$ then $z=0$.

From this one can calculate the coefficient $c_l$, giving 
\begin{equation}
c_{l}=\frac{c_{R}\overset{l}{\underset{k=2}{\prod }}R_{2k-1}^{2}\overset{m}{%
\underset{k=l}{\prod }}R_{2k}^{2}}{\overset{l-1}{\underset{k=1}{\prod }}%
R_{2k}^{2}\overset{m}{\underset{k=l+1}{\prod }}R_{2k-1}^{2}}\, ,
\end{equation}%
where the subscript on $c_R$ denotes the representation $R$ and not a
radius. Before taking the $R_{1}^{2}=0$ limit the expression has the
property that the coherent state wavefunction is normalized to zero at the
origin, because there is no black region there. The limit from $2m$ circles
to $2m-1$ circles is similar.

The overall normalization factor for each black ring has the property that: 
\begin{eqnarray}
\left\langle z|z\right\rangle _{l} &=&\frac{c_{l}\overset{l-1}{\underset{k=1}%
{\prod }}\left( R_{2k}^{2}-\left\vert z\right\vert ^{2}\right) \overset{m}{%
\underset{k=l+1}{\prod }}\left( R_{2k-1}^{2}-\left\vert z\right\vert
^{2}\right) (-\left\vert z\right\vert ^{2})}{\overset{l}{\underset{k=1}{%
\prod }}\left( R_{2k-1}^{2}-\left\vert z\right\vert ^{2}\right) \overset{m}{%
\underset{k=l}{\prod }}\left( R_{2k}^{2}-\left\vert z\right\vert ^{2}\right) 
} \\
&=&\frac{c_{R}\overset{l-1}{\underset{k=1}{\prod }}\left( 1-\frac{\left\vert
z\right\vert ^{2}}{R_{2k}^{2}}\right) \overset{m}{\underset{k=l+1}{\prod }}%
\left( 1-\frac{\left\vert z\right\vert ^{2}}{R_{2k-1}^{2}}\right) \frac{%
(-\left\vert z\right\vert ^{2})}{\left( R_{1}^{2}-\left\vert z\right\vert
^{2}\right) }}{\overset{l}{\underset{k=2}{\prod }}\left( 1-\frac{\left\vert
z\right\vert ^{2}}{R_{2k-1}^{2}}\right) \overset{m}{\underset{k=l}{\prod }}%
\left( 1-\frac{\left\vert z\right\vert ^{2}}{R_{2k}^{2}}\right) }\,.
\end{eqnarray}%
When one crosses from the $l^{th}$ ring to the $(l+1)^{th}$ ring, there is
only a change of factor

\begin{equation}
\frac{\left( 1-\frac{\left\vert z\right\vert ^{2}}{R_{2l+1}^{2}}\right) }{%
\left( 1-\frac{\left\vert z\right\vert ^{2}}{R_{2l}^{2}}\right) }\rightarrow 
\frac{\left( 1-\frac{\left\vert z\right\vert ^{2}}{R_{2l}^{2}}\right) }{%
\left( 1-\frac{\left\vert z\right\vert ^{2}}{R_{2l+1}^{2}}\right) }\, .
\end{equation}%
Similarly, when one crosses from the $l^{th}$ ring to the $(l-1)^{th}$ ring,
there is only a change of factor 
\begin{equation}
\frac{\left( 1-\frac{\left\vert z\right\vert ^{2}}{R_{2l-2}^{2}}\right) }{%
\left( 1-\frac{\left\vert z\right\vert ^{2}}{R_{2l-1}^{2}}\right) }%
\rightarrow \frac{\left( 1-\frac{\left\vert z\right\vert ^{2}}{R_{2l-1}^{2}}%
\right) }{\left( 1-\frac{\left\vert z\right\vert ^{2}}{R_{2l-2}^{2}}\right) }%
\, .
\end{equation}%
These are due to the sign changes of the contributions in $\left\vert
z\right\rangle $ from the two circles that one passes by.

The disappearance of the central circle leads to the limit 
\begin{equation}
\frac{(-\left\vert z\right\vert ^{2})}{\left( R_{1}^{2}-\left\vert
z\right\vert ^{2}\right) }\rightarrow 1\,,
\end{equation}%
which gives a good limit to the case of an odd number of circles - i.e. a
disk plus annuli configuration. This is also a good limit for $\left\vert
z\right\rangle $, since $-\frac{R_{1}^{2}}{R_{1}^{2}-\left\vert z\right\vert
^{2}}\rightarrow 0$.

A merging of two black rings when $R_{2j}^{2}\rightarrow R_{2j+1}^{2}$ is
also a natural and well behaved limit, as, e.g.%
\begin{equation}
\frac{\left( 1-\frac{\left\vert z\right\vert ^{2}}{R_{2j+1}^{2}}\right) }{%
\left( 1-\frac{\left\vert z\right\vert ^{2}}{R_{2j}^{2}}\right) }\rightarrow
1,\qquad \frac{\left( 1-\frac{\left\vert z\right\vert ^{2}}{R_{2j}^{2}}%
\right) }{\left( 1-\frac{\left\vert z\right\vert ^{2}}{R_{2j+1}^{2}}\right) }%
\rightarrow 1\, ,
\end{equation}%
so the formula reduces to the situation with one fewer black ring. This is
also a good limit for $\left\vert z\right\rangle $, since $%
c_{l}^{2j+1}\rightarrow -c_{l}^{2j}$, and contributions from the two merging
circles cancel.

The situation where the $l^{th}$ black ring is far away from others leads to
the following limit 
\begin{equation}
\left\langle z|z\right\rangle _{l}\rightarrow c_{l}\frac{(-\left\vert
z\right\vert ^{2})}{\left( R_{2l-1}^{2}-\left\vert z\right\vert ^{2}\right)
\left( R_{2l}^{2}-\left\vert z\right\vert ^{2}\right) }=\frac{%
c_{l}^{2l-1}R_{2l-1}^{2}}{R_{2l-1}^{2}-\left\vert z\right\vert ^{2}}+\frac{%
c_{l}^{2l}R_{2l}^{2}}{R_{2l}^{2}-\left\vert z\right\vert ^{2}}\,,
\end{equation}%
which reduces to the one black ring case, which is a form of decoupling
limit for the $l^{th}$ ring.

\vspace{1pt}The limit where we go from $2m$ circles to $2m-1$ circles takes
the following form 
\begin{equation}
R_{1}^{2}\rightarrow 0,\qquad V\rightarrow V\,.
\end{equation}%
In fact the 2-circle case can be viewed as a limit of the 3-circle case with 
$R_{1}=0$ leading to the $1/\bar{z}$ term.

\vspace{1pt}There is a further limit which would give the wavefunction in
complimentary regions nomarlizable on the white regions instead of the
black. Starting from the case of an even number (e.g. $2m$) of circles, and
sending the last circle to infinity, followed by switching the black and
white regions leads to

\begin{equation}
R_{2m}^{2}\rightarrow \infty ,\qquad V\rightarrow -V\, .
\end{equation}%
This procedure gives us coherent state wavefunction in white regions instead
of black, as it is complimentary to the wavefunction in the black regions.

\section{Dimension of the density matrix}

\label{appendix_ dimension_denstiy_matrix} The dimension of the
density matrix $D$ in section \ref{denstiy_matrix} is related to the entropy
of the superstar, $S$ via%
\begin{equation}
e^{S}=D\,.
\end{equation}%
\vspace{1pt}The superstar geometry corresponds to a grey disk on the $z$%
-plane with a larger radius and lower fermion filling density than the AdS
disk. It has the same $N$ as the AdS geometry but has a larger radius on the 
$z$-plane \cite{Lin:2004nb},\cite{Gubser:2004xx},\cite%
{Balasubramanian:2005mg}:%
\begin{eqnarray}
R_{star}^{2} &=&R_{AdS}^{2}(1+\frac{2\Delta }{N^{2}})=N+M\,,~~~~~\
R_{AdS}^{2}=N\,, \\
M &=&\frac{2\Delta }{N}\,.
\end{eqnarray}%
A single superstar may be viewed as a microcanonical ensemble with $N$ total
rows, and $M$ total columns, both fixed. In other words, we have 
\begin{equation}
N=\sum\limits_{j=1}^{m}N_{j},~~\ \ M=\sum\limits_{j=1}^{m}M_{j}\,.
\end{equation}%
The slope of such a Young tableau is, in the limit of large $M$ and $N$, 
\begin{equation}
q=\frac{M}{N}=\frac{2\Delta }{N^{2}}\,,
\end{equation}%
and the filling density on the grey disk with the larger radius, is $\frac{1%
}{1+q}=\frac{N}{N+M}$.

One can consider the entropy as the sum of entropies of each area element in
the phase space, each of which is subject to a Bernoulli distribution $%
\{u,1-u\}$. The sum is written in \cite{D'Errico:2007jm} (see also \cite%
{Balasubramanian:2007zt}) which is%
\begin{eqnarray}
S &=&-\frac{1}{2\pi \hbar }\int d^{2}x\left[ u\log u+(1-u)\log (1-u)\right]
\label{information_shannon} \\
&=&-N\int d\left\vert z\right\vert ^{2}\left[ u\log u+(1-u)\log (1-u)\right]
\,,
\end{eqnarray}%
where the area is quantized due to the quantization of the phase space%
\begin{equation}
\frac{1}{2\pi \hbar }\int d^{2}x=M+N=N\int d\left\vert z\right\vert
^{2},\quad \frac{R^{2}|_{there}}{2\hbar }=\frac{R^{2}|_{here}}{1/N}\,,
\label{phase_space_quantization}
\end{equation}%
(see the footnote on page 3). For a uniform distribution in the phase space, 
$u$ is a constant which is the filling fraction $\frac{N}{M+N}$, and we can
thus write the entropy as%
\begin{eqnarray}
S &=&(M+N)\left[ \frac{M}{M+N}\text{log}\left( \frac{M+N}{M}\right) +\frac{N%
}{M+N}\text{log}\left( \frac{M+N}{N}\right) \right]  \label{entropy_N_M} \\
&=&~f\left( \frac{N}{N+M}\right) \cdot \sqrt{\Delta }\,,
\label{entropy_dimension}
\end{eqnarray}%
where 
\begin{equation}
f\left( \frac{N}{N+M}\right) =\left( \sqrt{\frac{2M}{N}}+\sqrt{\frac{2N}{M}}%
\right) \left[ \frac{M}{M+N}\text{log}\left( \frac{M+N}{M}\right) +\frac{N}{%
M+N}\text{log}\left( \frac{M+N}{N}\right) \right] \,.
\end{equation}%
This agrees exactly with \cite{Balasubramanian:2005mg}, which is derived
from the partition function approach.

In the case of unrestricted $N$ and $M$, the Hardy-Ramanujan formula tells
us that \cite{Suryanarayana:2004ig},\cite{Balasubramanian:2005mg}%
\begin{equation}
S\sim \sqrt{\frac{2\pi ^{2}}{3}}\sqrt{\Delta }=\sqrt{\frac{\pi ^{2}}{3}MN}%
=2.5651\sqrt{\Delta }\, ,
\end{equation}%
on the other hand from (\ref{entropy_dimension}) 
\begin{equation}
S=f\left( \frac{N}{N+M}\right) \sqrt{\Delta }\leqslant 1.96052\sqrt{\Delta }%
\, ,
\end{equation}%
which is consistent with the Hardy-Ramanujan bound.

In the $M\gg N$ regime (\ref{entropy_dimension}) has the limit, 
\begin{equation}
S\sim N\left( \text{log}\frac{M}{N}+1\right)\, .  \label{entropy_M_large}
\end{equation}
This formula is similar to the case of the type IIA geometry with $N$ units
of D2-brane flux and $K$ units of NS5-brane flux on the torus \cite%
{Shieh:2007xn}.

Another method to obtain this order of magnitude estimate for the entropy in
the superstar geometry is by using Schur's theorem to find the number of
partitions of $\Delta =$ $\frac{MN}{2}~$with $N$ maximum rows. If $D$
denotes the number of possible sets $n_{1},...,n_{N}$ which satisfy the
relation $\Delta =\sum_{i=1}^{N}r_{i}n_{i}$ and are non-negative, and $%
\{r_{i}\}~$are relatively prime, then 
\begin{equation}
D\sim {\frac{\Delta ^{N-1}}{(N-1)!}}{\frac{1}{r_{1}\dots r_{N}}}\,,
\label{schur_scaling}
\end{equation}%
in the large $\Delta $ limit. This should give an order of magnitude
estimate, with the distinction coming from the fact that the set $\{r_{i}\}$
may not be relatively prime. If we let $r_{k}=k$, which is the case we are
interested in, we find

\begin{equation}
\log D\sim N\left(\log {\frac{\Delta }{N^{2}}}+2\right)\, ,
\label{dimension_Schur_scaling}
\end{equation}%
where we have used Stirling's approximation for the logarithm of the
factorial in the large $N$ limit. We see that (\ref{dimension_Schur_scaling}%
) agrees with (\ref{entropy_M_large}) in the leading order.

\section{Procedure for computing the density matrix}
\label{appendix_computational_density_matrix}

\begin{figure}[h]
\begin{center}
\includegraphics[width=7cm]{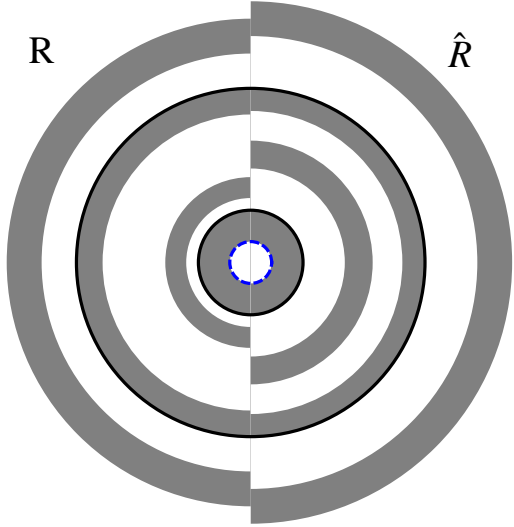}
\end{center}
\caption{{\protect\small Two representations $R$ and $\hat{R}$ given by
their half plane disk diagrams. The black solid and blued dashed circles
emphasize the coincident outer and inner rings.}}
\label{fig.example}
\end{figure}

In this section we will give a detailed explanation
as to how to calculate the reduced density matrix given a set of Young
tableaux. We start with two representations shown in figure \ref{fig.example}%
. The two ring configurations for the representations $R$ and $\hat{R}$ are
drawn only in their half planes so that we may see the coincident outer
rings (marked with black lines) and coincident inner rings (marked with
dashed blue lines).\ \vspace{1pt} 
We may draw a more detailed picture of these two representations by simply
plotting the radii in a linear fashion, as shown in figure \ref%
{fig.coincidentrings}. In this figure we have also labelled the $a_{k}$ and $%
b_{k}$ oscillators for the $R$ representation along with the eight radii $%
R_{1},...,R_{8}$ and similarly the $\hat{a}_{k},\hat{b}_{k}$ and $\hat{R}%
_{i} $ for the $\hat{R}$ representation.

We see that the ring radii are coincident on two outer rings and one inner
ring (the innermost). Note that the actual values of the radii are
unimportant for this example. We simply aim to show the calculational
procedure. 
\begin{figure}[h]
\begin{center}
\includegraphics[width=15cm]{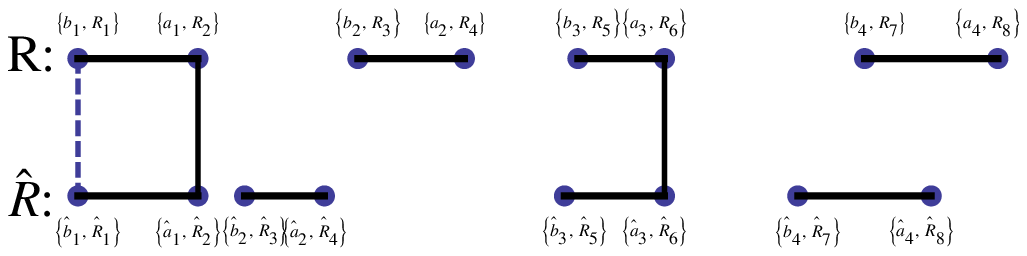}
\end{center}
\caption{{\protect\small The two representations $R$ and $\hat{R}$ compared
in a purely radial diagram, showing their coincident inner and outer edges
along with their $a$, $b$ and $R_{i}$ labels.}}
\label{fig.coincidentrings}
\end{figure}

We first write the coherent state wave function for the first ring structure
(which has four rings) as: 
\begin{equation}
\left\vert z\right\rangle =\left\vert z\right\rangle _{1}\theta
_{R_{1},R_{2}}+\left\vert z\right\rangle _{2}\theta
_{R_{3},R_{4}}+\left\vert z\right\rangle _{3}\theta
_{R_{5},R_{6}}+\left\vert z\right\rangle _{4}\theta _{R_{7},R_{8}}\, ,
\end{equation}%
and for the second 
\begin{equation}
\left\vert \hat{z}\right\rangle =\left\vert \hat{z}\right\rangle _{1}\theta
_{\hat{R}_{1},\hat{R}_{2}}+\left\vert \hat{z}\right\rangle _{2}\theta _{\hat{%
R}_{3},\hat{R}_{4}}+\left\vert \hat{z}\right\rangle _{3}\theta _{\hat{R}_{5},%
\hat{R}_{6}}+\left\vert \hat{z}\right\rangle _{4}\theta _{\hat{R}_{7},\hat{R}%
_{8}}\, ,
\end{equation}%
where each coherent state wavefunction $\left\vert \hat{z}\right\rangle _{l}$
has support only on the $l^{th}$ ring $\theta_{xy}=1$ for $x<z<y$ and zero
otherwise. Note that these are coherent states, though we can multiply the
whole state by an arbitrary constant and it will still be a coherent state.
We set this arbitrary normalization such that the inner product of the state
with itself, integrated over the whole space gives $\frac{1}{D}$ where $D$
is the total number of representations in the ensemble. 
\begin{equation}
\int d^2 z\left\langle {z}|z\right\rangle=\int d^2 z\left\langle \hat{z}|%
\hat{z}\right\rangle=1\, .
\end{equation}
We can write out the coherent state wavefunction in full, in terms of the $%
f^{\prime}s$ defined in (\ref{eq.fs}) 
\begin{equation}
\left\vert z\right\rangle =\overset{4}{\underset{l=1}{\sum }}\left\{ \overset%
{\infty }{\underset{n=0}{\sum }}\overset{l}{\underset{k=1}{\sum }}%
f_{l}^{2k-1}\left( \frac{{R}_{2k-1}}{z}\right) ^{n}\left\vert
-n,b_{k}\right\rangle +\overset{\infty }{\underset{n=1}{\sum }}\overset{4}{%
\underset{k=l}{\sum }}f_{l}^{2k}\left( \frac{z}{{R}_{2k}}\right)
^{n}\left\vert n,a_{k}\right\rangle \right\} \theta _{{R}_{2l-1},\hat{R}%
_{2l}}\, ,
\end{equation}

\begin{equation}
\left\vert \hat{z}\right\rangle =\overset{4}{\underset{{\hat{l}}=1}{\sum }}%
\left\{ \overset{\infty }{\underset{n=0}{\sum }}\overset{\hat{l}}{\underset{%
\hat{k}=1}{\sum }}{\hat{f}}_{\hat{l}}^{2\hat{k}-1}\left( \frac{\hat{R}_{2%
\hat{k}-1}}{z}\right) ^{n}\left\vert -n,b_{\hat{k}}\right\rangle +\overset{%
\infty }{\underset{n=1}{\sum }}\overset{4}{\underset{\hat{k}=\hat{l}}{\sum }}%
{\hat{f}}_{\hat{l}}^{2\hat{k}}\left( \frac{z}{{\hat{R}}_{2\hat{k}}}\right)
^{n}\left\vert n,a_{\hat{k}}\right\rangle \right\} \theta _{\hat{R}_{2\hat{l}%
-1},\hat{R}_{2\hat{l}}}\, .
\end{equation}
Note that the fact that there are four rings in both representations is a
coincidence - in many cases one looks at the overlap between representations
with different numbers of rings.

Now let's look at the inner product of these two objects. There is one clear
simplification we can make. This is that there are never any cross products
between the $a$'s and the $b$'s, so we can write:

\begin{equation}
\left\langle \hat{z}|z\right\rangle =\left\langle \hat{z}|z\right\rangle
_{a}+\left\langle \hat{z}|z\right\rangle _{b}\,.  \label{contributions_a_b}
\end{equation}%
Looking first at the $a$ contribution in (\ref{contributions_a_b}) we see
that we have the following sums: 
\begin{eqnarray}
\left\langle z,\hat{R}|z,R\right\rangle _{a} &=&\overset{4}{\underset{l=1}{%
\sum }}\overset{4}{\underset{\hat{l}=1}{\sum }}\overset{\infty }{\underset{%
n=1}{\sum }}\overset{\infty }{\underset{m=1}{\sum }}\overset{4}{\underset{k=l%
}{\sum }}\overset{4}{\underset{\hat{k}=\hat{l}}{\sum }}{\hat{f}}_{\hat{l}}^{2%
\hat{k}}{f}_{l}^{2k}\left( \frac{z}{R_{2k}}\right) ^{n}\left( \frac{%
\overline{z}}{{\hat{R}}_{2\hat{k}}}\right) ^{m}<m,a_{\hat{k}}|n,a_{k}>\theta
_{R_{2l-1},R_{2l}}\theta _{\hat{R}_{2\hat{l}-1},\hat{R}_{2\hat{l}}}  \notag
\\
&=&\overset{4}{\underset{l=1}{\sum }}\overset{4}{\underset{\hat{l}=1}{\sum }}%
\overset{\infty }{\underset{n=1}{\sum }}\overset{\infty }{\underset{m=1}{%
\sum }}\overset{4}{\underset{k=l}{\sum }}\overset{4}{\underset{\hat{k}=\hat{l%
}}{\sum }}{\hat{f}}_{\hat{l}}^{2\hat{k}}{f}_{l}^{2k}\left( \frac{z}{R_{2k}}%
\right) ^{n}\left( \frac{\overline{z}}{{\hat{R}}_{2\hat{k}}}\right)
^{m}\delta _{mn}\delta _{\hat{R}_{2\hat{k}}R_{2k}}\theta
_{R_{2l-1},R_{2l}}\theta _{\hat{R}_{2\hat{l}-1},\hat{R}_{2\hat{l}}}  \notag
\\
&=&\overset{4}{\underset{l=1}{\sum }}\overset{4}{\underset{\hat{l}=1}{\sum }}%
\overset{\infty }{\underset{n=1}{\sum }}\overset{4}{\underset{k=l}{\sum }}%
\overset{4}{\underset{\hat{k}=\hat{l}}{\sum }}{\hat{f}}_{\hat{l}}^{2\hat{k}}{%
f}_{l}^{2k}\left( \frac{\left\vert z\right\vert }{R_{2k}}\right) ^{2n}\delta
_{\hat{R}_{2\hat{k}}R_{2k}}\theta _{R_{2l-1},R_{2l}}\theta _{\hat{R}_{2\hat{l%
}-1},\hat{R}_{2\hat{l}}}  \notag \\
&=&\overset{4}{\underset{l=1}{\sum }}\overset{4}{\underset{\hat{l}=1}{\sum }}%
\overset{4}{\underset{k=l}{\sum }}\overset{4}{\underset{\hat{k}=\hat{l}}{%
\sum }}{\hat{f}}_{\hat{l}}^{2\hat{k}}{f}_{l}^{2k}\left( \frac{\left\vert
z\right\vert ^{2}}{R_{2k}^{2}-\left\vert z\right\vert ^{2}}\right) \delta _{%
\hat{R}_{2\hat{k}}R_{2k}}\theta _{R_{2l-1},R_{2l}}\theta _{\hat{R}_{2\hat{l}%
-1},\hat{R}_{2\hat{l}}}\,.  \notag \\
&&
\end{eqnarray}%
There are two positions where the outer rings overlap, for $l=\hat{l}=1$ and 
$l=\hat{l}=3$. Note that it does not have to be the case that they are on
the same rings. From this observation we get the following two terms
contributing:

\begin{eqnarray}
\left\langle z,\hat{R}|z,R\right\rangle _{a} &=&{\hat{f}}_{1}^{2}{f}%
_{1}^{2}\left( \frac{\left\vert z\right\vert ^{2}}{R_{2}^{2}-\left\vert
z\right\vert ^{2}}\right) \theta _{R_{1},R_{2}}\theta _{\hat{R}_{1},\hat{R}%
_{2}}  \notag \\
&&+\left( \frac{\left\vert z\right\vert ^{2}}{R_{6}^{2}-\left\vert
z\right\vert ^{2}}\right) \left( {\hat{f}}_{3}^{6}{f}_{3}^{6}\theta
_{R_{5},R_{6}}\theta _{\hat{R}_{5},\hat{R}_{6}}+{\hat{f}}_{1}^{6}{f}%
_{1}^{6}\theta _{R_{1},R_{2}}\theta _{\hat{R}_{1},\hat{R}_{2}}\right)\, .
\end{eqnarray}

The first term includes a contribution from only one ring, whereas the
second term, which comes from coincident $a$'s on the third ring gives
contributions from all the overlapping regions of all inward rings from that
point.

We can write a similar expression for the $b$'s and we find that the process
is very similar except that the contribution comes from all rings out from
the coincident radii. This time there is clearly no contribution from the
second rings as they do not overlap at all, so there are again just three
contributing terms here 
\begin{equation}
\left\langle z,\hat{R}|z,R\right\rangle _{b}=\left( \frac{\left\vert
z\right\vert ^{2}}{\left\vert z\right\vert ^{2}-R_{1}^{2}}\right) \left( 
\hat{f}_{1}^{1}f_{1}^{1}\theta _{R_{1}R_{2}}\theta _{\hat{R}_{1}\hat{R}_{2}}+%
\hat{f}_{3}^{1}f_{3}^{1}\theta _{R_{5}R_{6}}\theta _{\hat{R}_{5}\hat{R}_{6}}+%
\hat{f}_{4}^{1}f_{4}^{1}\theta _{R_{7}R_{8}}\theta _{\hat{R}_{7}\hat{R}%
_{8}}\right) \,.
\end{equation}%
Note that it is just a coincidence that the radii have the same label in the
products of $\theta $'s - they may come from overlaps of very different ring
numbers in more complex examples. Because of the particular overlaps we see
in the above diagram, we can simplify the $\theta $, to get: 
\begin{equation}
\left\langle z,\hat{R}|z,R\right\rangle _{a}=\left( \frac{\left\vert
z\right\vert ^{2}}{R_{2}^{2}-\left\vert z\right\vert ^{2}}\right) {\hat{f}}%
_{1}^{2}{f}_{1}^{2}\theta _{R_{1},R_{2}}+\left( \frac{\left\vert
z\right\vert ^{2}}{R_{6}^{2}-\left\vert z\right\vert ^{2}}\right) \left( {%
\hat{f}}_{3}^{6}{f}_{3}^{6}\theta _{R_{5},R_{6}}+{\hat{f}}_{1}^{6}{f}%
_{1}^{6}\theta _{R_{1},R_{2}}\right) \,,
\end{equation}%
\begin{equation}
\left\langle z,\hat{R}|z,R\right\rangle _{b}=\left( \frac{\left\vert
z\right\vert ^{2}}{\left\vert z\right\vert ^{2}-R_{1}^{2}}\right) \left( 
\hat{f}_{1}^{1}f_{1}^{1}\theta _{R_{1}R_{2}}+\hat{f}_{3}^{1}f_{3}^{1}\theta
_{R_{5}R_{6}}+\hat{f}_{4}^{1}f_{4}^{1}\theta _{R_{7}\hat{R}_{8}}\right) \,.
\end{equation}%
We will then need to take the integral over the complex $z$ plane. The
factors of $2\pi $ are not important because they will be taken care of when
we normalize the diagonal of the reduced density matrix to $\frac{1}{D}$
where $D$ is the number of diagrams contributing. The reduced density matrix
therefore between these two terms, up to normalization is given by: 
\begin{eqnarray}
\left\langle \hat{R}|R\right\rangle &=&\int d\left\vert z\right\vert \left( 
\frac{\left\vert z\right\vert ^{3}}{R_{2}^{2}-\left\vert z\right\vert ^{2}}%
\right) {\hat{f}}_{1}^{2}{f}_{1}^{2}\theta _{R_{1},R_{2}}+\left( \frac{%
\left\vert z\right\vert ^{3}}{R_{6}^{2}-\left\vert z\right\vert ^{2}}\right)
\left( {\hat{f}}_{3}^{6}{f}_{3}^{6}\theta _{R_{5},R_{6}}+{\hat{f}}_{1}^{6}{f}%
_{1}^{6}\theta _{R_{1},R_{2}}\right)  \notag \\
&&+\left( \frac{\left\vert z\right\vert ^{3}}{\left\vert z\right\vert
^{2}-R_{1}^{2}}\right) \left( \hat{f}_{1}^{1}f_{1}^{1}\theta _{R_{1}R_{2}}+%
\hat{f}_{3}^{1}f_{3}^{1}\theta _{R_{5}R_{6}}+\hat{f}_{4}^{1}f_{4}^{1}\theta
_{R_{7}\hat{R}_{8}}\right) \,.
\end{eqnarray}%
In fact this integral is divergent, but there is a very natural cutoff given
by the fact that the phase space is discrete. The integration of the area in
the phase space is really the sum of area elements in the phase space, i.e. $%
d(\left\vert z\right\vert ^{2})=\delta (R^{2})$. Given that the $R^{2}$ are
quantized in units of $\frac{1}{N}$, we subtract (or add) $\frac{1}{N}$ from the upper (lower)
bound given by the $R_{2k-1}$ ($R_{2k}$), giving: 
\begin{eqnarray}
\left\langle \hat{R}|R\right\rangle &=&{\hat{f}}_{1}^{2}{f}_{1}^{2}\left.
\left( -\frac{R_{2}^{2}}{2}\log \left( R_{2}^{2}-\left\vert z\right\vert
^{2}\right) -\frac{\left\vert z\right\vert ^{2}}{2}\right) \right\vert
_{R_{1}}^{\sqrt{R_{2}^{2}-\frac{1}{N}}}  \notag \\
&&+{\hat{f}}_{3}^{6}{f}_{3}^{6}\left. \left( -\frac{R_{6}^{2}}{2}\log \left(
R_{6}^{2}-\left\vert z\right\vert ^{2}\right) -\frac{\left\vert z\right\vert
^{2}}{2}\right) \right\vert _{R_{5}}^{\sqrt{R_{6}^{2}-\frac{1}{N}}}+{\hat{f}}_{1}^{6}{f%
}_{1}^{6}\left. \left( -\frac{R_{6}^{2}}{2}\log \left( R_{6}^{2}-\left\vert
z\right\vert ^{2}\right) -\frac{\left\vert z\right\vert ^{2}}{2}\right)
\right\vert _{R_{1}}^{R_{2}}  \notag \\
&&+\hat{f}_{1}^{1}f_{1}^{1}\left. \left( \frac{R_{1}^{2}}{2}\log \left(
R_{1}^{2}-\left\vert z\right\vert ^{2}\right) +\frac{\left\vert z\right\vert
^{2}}{2}\right) \right\vert _{\sqrt{R_{1}^{2}+\frac{1}{N}}}^{R_{2}}+\hat{f}%
_{3}^{1}f_{3}^{1}\left. \left( \frac{R_{1}^{2}}{2}\log \left(
R_{1}^{2}-\left\vert z\right\vert ^{2}\right) +\frac{\left\vert z\right\vert
^{2}}{2}\right) \right\vert _{R_{5}}^{R_{6}}  \notag \\
&&+\hat{f}_{4}^{1}f_{4}^{1}\left. \left( \frac{R_{1}^{2}}{2}\log \left(
R_{1}^{2}-\left\vert z\right\vert ^{2}\right) +\frac{\left\vert z\right\vert
^{2}}{2}\right) \right\vert _{R_{7}}^{R_{8}}\,.
\end{eqnarray}%
This whole expression needs to be correctly normalized such that $%
\left\langle R|R\right\rangle $ and $\left\langle \hat{R}|\hat{R}%
\right\rangle $ are $\frac{1}{D}$. Note that for small $N$ and $M$ one
should really apply a summation rather than an integration at this point.
This distinction will not be important in the large $M,N$ limit. There are
however subtleties when we fix, for instance large $M$ and small $N$ due to
coincident thin rings at very large radii. For small $N$, these dominate
when one uses the summation. If one considers an ensemble in which the total 
$M$ is allowed to vary slightly, such outer ring coincidences are rare and
this issue becomes unimportant.

\begin{figure}[h]
\begin{center}
\includegraphics[width=9.5cm]{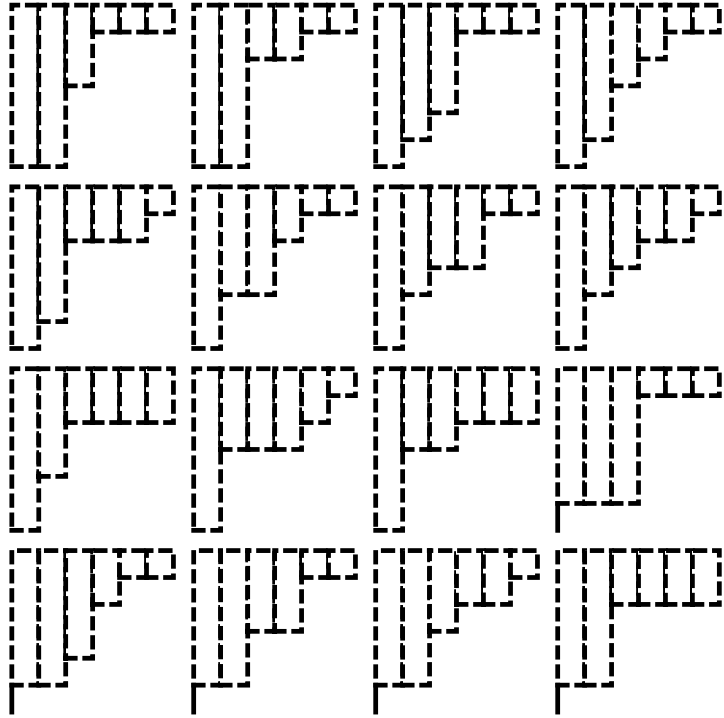}
\end{center}
\caption{{\protect\small 16 examples of the 29 Young tableaux making up the ensemble of $%
M=N=6$, $\Delta=18$.}}
\label{fig.66diagrams}
\end{figure}

\begin{figure}[h]
\begin{center}
\includegraphics[width=10cm]{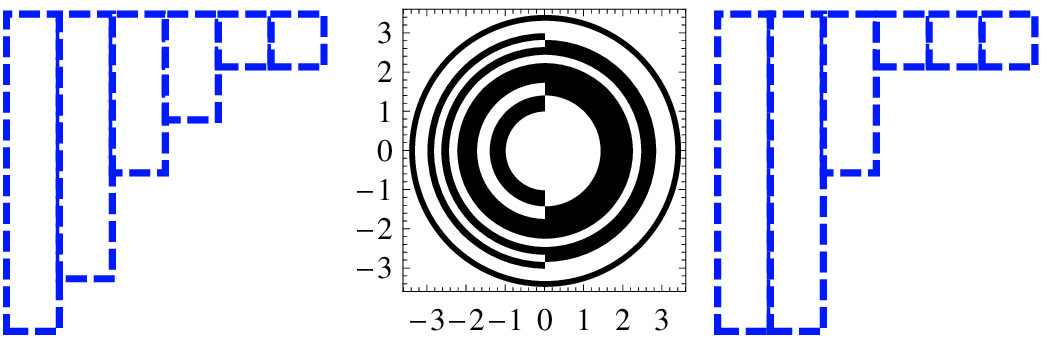}
\end{center}
\caption{{\protect\small The $1^{st}$ and $4^{th}$ diagrams from figure 
\protect\ref{fig.66diagrams} compared. The (1,4) reduced density element is $%
\frac{1}{D}0.12$ showing small overlap in the representations.}}
\label{fig.bluediags}
\end{figure}

\begin{figure}[h]
\begin{center}
\includegraphics[width=10cm]{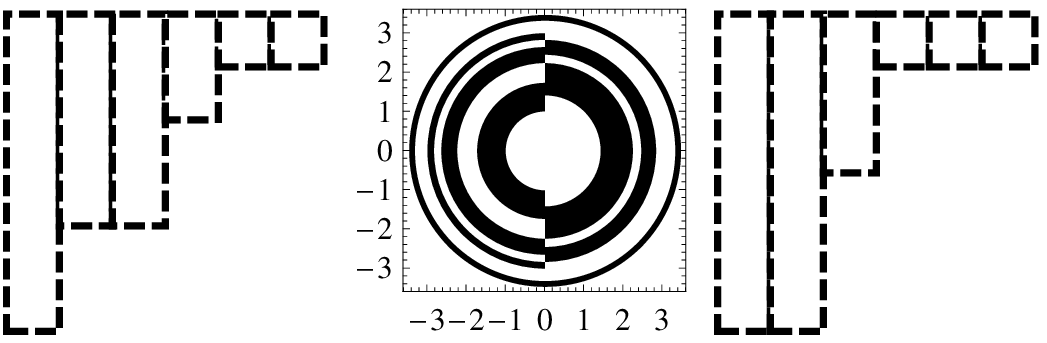}
\end{center}
\caption{{\protect\small The $1^{st}$ and $6^{th}$ diagrams from figure 
\protect\ref{fig.66diagrams} compared. The (1,6) reduced density element is $%
\frac{1}{D}0.0014$ showing almost no overlap in the representations.}}
\label{fig.blackdiags}
\end{figure}

Having given an abstract example, we turn to a concrete calculation. In this
case we look at the case for $M=N=6$ and $\Delta=\frac{M N}{2}=18$. In this
case there are a total of 29 Young tableaux which make up the ensemble. Some examples of these are shown in figure \ref{fig.66diagrams}. The five diagrams with solid lines
extending from the base are those which have central black disks within the
annuli. 
We can calculate the reduced density matrix from these diagrams using the
previously described algorithm and find the graph shown in figure \ref%
{fig.pl6}. We concentrate on 2 points in the 29 by 29 matrix which
correspond to the mixing between different representations. The two
pairs of Young tableaux and disk configurations in the $z$-plane are shown
in figures \ref{fig.bluediags} and \ref{fig.blackdiags}
where the values of these reduced density matrix elements are $\frac{1}{D}0.12$ and $\frac{1}{D}0.0014$ respectively, where $D=29$
is the dimension of the reduced density matrix. The fact that these
off-diagonal values are non-zero shows that the entropy will be reduced from
the maximal value of $\log(D)$. The mixing of representations via the
breaking of supersymmetry leads directly to these non-zero off-diagonals. 


From the above discussion we can summarize the computational procedure for
calculating the entropy of any chosen ensemble. The steps in the procedure
are as follows:


1. Choose the values of $M$, $N$ and $\Delta $. Some of these may or may not
be constrained depending on the limits of interest.

2. Given the constraints on $M$, $N$ and $\Delta $ calculate all of the
Young tableaux (including inner disk contributions) for these parameters. In
general the number of diagrams, $D$, goes roughly as $e^{\sqrt{\Delta }}$
though the constraints on $N$ and $M$ can lower this number considerably.

3. Look first at the diagonal elements of the reduced density matrix. There
is an overall normalization of every $\left\vert z\right\rangle $ which is
fixed by requiring that each diagonal element of the reduced density matrix
should be $\frac{1}{D}$. These normalizations are stored and are later used
to weight the off-diagonal elements appropriately.

4. Study the off-diagonal elements by cycling through all pairs of
representations.

5. Given two representations we first look for all coincident inner and
outer edges which will contribute to the total reduced density matrix
element. Calculate individually the contributions from the inner and outer
edges, integrating over the appropriate overlapping regions, summing to give
the total and dividing by the weighting factors calculated from the diagonal
elements.

6. Calculate $S=-Tr(\rho \log {\rho })$ by diagonalizing $\rho$.


\section{Derivation of the BMN spectrum}
\label{appendix_BMN_states}

We include here an overview of the calculation of the BMN state
using a series of recursion relations obtained from the energy eigenstate
condition. The equation we discuss here is (\ref%
{generating_recursive_equations}). This equation was obtained in \cite%
{Chen:2007gh} but the details of the calculation were not given there. We
think that it might be useful to the reader to include more detailed
intermediate steps in this appendix. Our results parallel theirs
when the parameter $\frac{N}{N_{l}}E$ is replaced by $E$.

The constraint that the state is an eigenfunction of the Hamiltonian relates
triples of $f_{n}^{\prime }s$. We are able to solve the system iteratively
by starting from two $f_{n}^{\prime }s$ which we chose to be $f_0 $ and $f_1$%
. The procedure has to be done in parts because different ranges of $n$ have
distinct recursion relations which cannot be solved using a generating
function formalism. We start by looking at the modes from $n=2$ to $J-3$ (to
be sure that we are not taking in any of the behaviour for the $J^{th}$ mode
which has a slightly different recursion relation due to the factors of $%
\gamma $ coming both from the prefactor and the action with $H$. We also
define $\frac{N}{N_{l}}E={\hat{E}~}$in (\ref{generating_recursive_equations}%
).

We use a method of generating functions which can be defined from a general
recursion relation. For the case where we have relations between triples of
coefficients in our recursion relation we have an equation of the form: 
\begin{equation}
af_{n+1}=bf_{n}+cf_{n-1}\, ,  \label{recursion_relation}
\end{equation}%
and define the generating function which we would like to solve for as 
\begin{equation}
B=\sum_{n=1}^{\infty }f_{n}x^{n}\, ,
\end{equation}%
we can multiply the above recursion relation (\ref{recursion_relation}) by $%
x^{n}$ and make the sum: 
\begin{equation}
\frac{1}{x}\sum_{n=2}^{\infty }af_{n}x^{n}=bB+x\sum_{n=0}^{\infty
}cf_{n}x^{n}\, ,
\end{equation}%
leading to the relation: 
\begin{equation}
\frac{1}{x}(aB-af_{1}x)=bB+x(cB+cf_{0})\, .
\end{equation}%
Solving for $B$ we have: 
\begin{equation}
B=\frac{xcf_{0}+af_{1}}{\frac{1}{x}a-b-cx}\, .
\end{equation}%
In the case of the energy eigenfunction equation, concretely we have that: 
\begin{equation}
B=-\frac{2x(\gamma +1)\lambda \left( x\sqrt{\frac{\gamma }{\gamma +1}}%
f_{0}-f_{1}\right) }{2(\gamma +1)\lambda (x-1)^{2}+{\hat{E}}x}\, ,
\label{B_function}
\end{equation}%
which we can write in the form of a summation by re-summing the geometric
series, such that: 
\begin{equation}
B=\frac{x\left( x\sqrt{\frac{\gamma }{\gamma +1}}f_{0}-f_{1}\right) }{%
r_{-}^{2}-1}\sum_{n=0}^{\infty }\left( \frac{1}{r_{-}^{n}}%
-r_{-}^{n+2}\right) x^{n}\, ,
\end{equation}%
where 
\begin{equation}
r_{-}=\frac{4\lambda (\gamma +1)+\sqrt{{\hat{E}}}\sqrt{{\hat{E}}-8\lambda
(\gamma +1)}-{\hat{E}}}{4\lambda (\gamma +1)}\, ,
\end{equation}%
By solving the recursion relations up to, for instance $n=(J-2)$ we can make
sure that this gives the correct closed form for the recursion relation. To
calculate $f_n$ one simply calculates the coefficient of the $x^n$ term. Now
we calculate $f_{J-1}$ up to $f_{J+1}$ in terms of $f_{0}$ and $f_{1}$ using
the recursion relation calculated and the generating function to take us to $%
f_{J-2}$, and find that: 
\begin{eqnarray}
f_{J+1} &=&\frac{r_{+}^{-J-1}\left( \gamma \left(
r_{+}^{2}-r_{+}^{2J}\right) \left( \left( \gamma _{+}+r_{+}\left( \gamma
_{+}r_{+}-1\right) \right) f_{0}+\gamma _{+}r_{+}f_{1}\right) \right) }{%
\gamma ^{2}\left( r_{+}^{2}-1\right) }  \notag \\
&+&\frac{r_{+}^{-J-1}\left( \sqrt{\gamma \gamma _{+}}\left( \gamma \left(
r_{+}^{2J}-r_{+}^{4}\right) f_{0}+r_{+}\left( r_{+}^{2J}-1\right) \left(
\gamma _{+}+r_{+}\left( \gamma _{+}r_{+}-1\right) \right) f_{1}\right)
\right) }{\gamma ^{2}\left( r_{+}^{2}-1\right)}\, ,  \notag \\
\end{eqnarray}%
where $r_{+}=\frac{1}{r_{-}}$ and $\gamma _{+}=\gamma +1$. Now we perform
the same procedure as before with a new generating functional, but going
from $n=J+2$ to $\infty $. This gives us the asymptotic behaviour for $f_{n}$
which we use to fix $f_{1}$. The generating functional, the equivalent of
equation (\ref{B_function}), is given by 
\begin{equation}
C=-\frac{2x^{J+1}\lambda \left( xf_{J}\gamma _{+}-\gamma f_{J+1}\right) }{%
\lambda (2x^{2}\gamma _{+}+\gamma (2-4x)-2x)+{\hat{E}}x}\, ,
\end{equation}%
from which we calculate that the $n^{th}$$>J^{th}$ term is given by: 
\begin{equation}
f_{n}=-\frac{\left( \gamma R_{+}\right) {}^{-n}\left( R_{+}\gamma
_{+}\right) {}^{-J}\left( \gamma ^{n}R_{+}^{2J+1}\left( \gamma
f_{J+1}-R_{+}\gamma _{+}f_{J}\right) \gamma _{+}^{J}+\gamma
^{J+1}R_{+}^{2n}\left( f_{J}-R_{+}f_{J+1}\right) \gamma _{+}^{n}\right) }{%
R_{+}^{2}\gamma _{+}-\gamma }\, ,
\end{equation}%
where 
\begin{equation}
R_{+}=\frac{2\lambda \left( \gamma +\gamma _{+}\right) -\sqrt{4\left(
\lambda ^{2}-{\hat{E}}\lambda \left( \gamma +\gamma _{+}\right) \right) +{%
\hat{E}}^{2}}-{\hat{E}}}{4\lambda \gamma _{+}}\, .
\end{equation}%
We want the eigenstate to be normalizable which puts a constraint on the $%
n\rightarrow \infty $ modes and leads to the constraint that: 
\begin{equation}
\gamma f_{J+1}-R_{+}\gamma _{+}f_{J}=0\, .
\end{equation}%
This is only possible for a particular value of $f_{1}$ and $f_{0}$. We have
the freedom, by the linearity of the eigenvalue equation, to set $f_{0}=1$
and so we are left with a constraint on $f_{1}$ which reads: 
\begin{equation}
f_{1}=\frac{\gamma \left( \left( r_{-}^{2J}-1\right) \gamma
_{+}r_{-}^{2}+\left( r_{-}^{2}-r_{-}^{2J}\right) \left( \gamma
_{+}R_{+}+1\right) r_{-}+\left( r_{-}^{2J}-r_{-}^{4}\right) \left( \gamma
_{+}-\sqrt{\gamma \gamma _{+}}\right) \right) }{r_{-}\sqrt{\gamma \gamma _{+}%
}\left( -r_{-}\left( r_{-}^{2J}-1\right) \left( \gamma _{+}R_{+}+1\right)
+\left( r_{-}^{2J+2}-1\right) \gamma _{+}+\left( r_{-}^{2J}-r_{-}^{2}\right)
\left( \gamma _{+}-\sqrt{\gamma \gamma _{+}}\right) \right) }\, .
\end{equation}%
We can check that this is correct by performing the full recursion order by
order in Mathematica up to a high value of $n$ and checking that the series
really does converge for $f_{1}$ of this form. Indeed the system is
extremely sensitive to tiny changes in $f_{1}$ and the above solution gives
precise convergence.

Now we must perform the same procedure but working to $n\rightarrow -\infty $
which leads to a constraint on ${\hat{E}}$ itself. Here using the recurrence
relation for negative $n$'s we get the following generating functional: 
\begin{eqnarray}
A &=&\sum_{n=0}^{\infty }f_{-n}x^{n}  \notag \\
&=&-\frac{x\left( -\gamma ^{2}\rho _{+}^{3}-\gamma \rho _{+}^{3}+\gamma
^{2}\rho _{+}\right) \left( -\gamma f_{-1}x^{2}-f_{-1}x^{2}+\gamma
f_{-2}x\right) }{\gamma (\gamma +1)\left( \rho _{+}-x\right) \left( x\rho
_{+}\gamma -\gamma +x\rho _{+}\right) \left( \frac{\gamma }{\gamma +1}-\rho
_{+}^{2}\right) }\, ,
\end{eqnarray}%
where: 
\begin{equation}
\rho =\frac{4\gamma \lambda +2\lambda -\sqrt{4\lambda ^{2}-8{\hat{E}}\gamma
\lambda -4{\hat{E}}\lambda +{\hat{E}}^{2}}-{\hat{E}}}{4(\gamma \lambda
+\lambda )}\, .
\end{equation}%
From this we can pull out the $n^{th}$ coefficient in the series expansion
to find the closed form value for $f_{-n}$. Stepping through the $n=-2$ and $%
n=-1$ terms from the $n=0$ and $n=1$ we can write this in terms of $f_{1}$
which has been fixed above. At this stage we have a closed form for $f_{-n}$
in terms of ${\hat{E}}$, $J$ and $n$. This means that we can look at the
asymptotics $n\rightarrow \infty $ and ask for a constraint on ${\hat{E}}$
such that the series is also convergent in this limit. Numerical studies
show that the eigenvalues, ${\hat{E}}$, coming from this constraint give
precisely the same values as those from equation (32) in \cite{Chen:2007gh}.

\begin{figure}[h]
\begin{center}
\includegraphics[width=8cm]{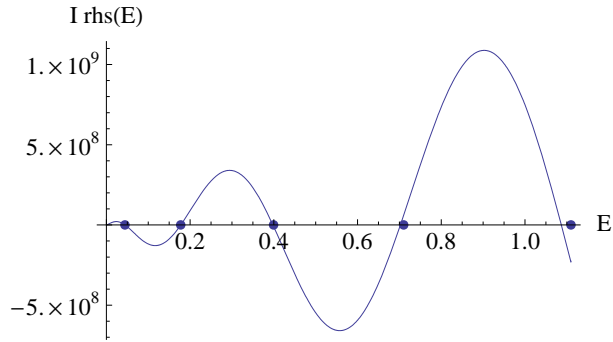}
\end{center}
\caption{{\protect\small Numerical calculation of the eigenvalues of the two
site lattice Hamiltonian.}}
\label{fig.rhs}
\end{figure}

The numerical method is performed as follows. By taking the equation for ${%
\hat{E},}$ similar to equation (32) in \cite{Chen:2007gh}, and plotting the
right hand side as a function of ${\hat{E}}$ and finding the points that it
crosses $0$ we are able to find the eigenvalues. In figure \ref{fig.rhs} we
plot $I=\sqrt{-1}$ times the right hand side of the equation versus ${\hat{E}%
}$ (solid line). The dots mark the points: 
\begin{equation}
{\hat{E}}_{\hat n}=(1+\gamma )\frac{2\pi ^{2}{\hat n}^{2}\lambda }{J^{2}}\left( 1-\frac{%
2+4\gamma }{J}\right)\, ,
\end{equation}%
\begin{equation}
E_{\hat n}=\frac{C_{a_{l}}}{N}\frac{2\pi ^{2}{\hat n}^{2}\lambda }{J^{2}}\left( 1-\frac{%
2+4\gamma }{J}\right)\, .
\end{equation}%
As discussed in section \ref{BMN_states}, one keeps only $\hat n=2n$ states as the physical BMN states, in agreement with \cite{Chen:2007gh}. Figure \ref{fig.rhs} is plotted for $J=30,\lambda =2,\gamma =\frac{1}{10}$. Note that one only expects agreement for the first few points for finite $J$
due to the number of solutions to the transcendental equation.

\end{document}